\documentclass[12pt]{article}
\usepackage{amsmath,amssymb,array,calc,rotating,epsfig,psfrag,amscd, cite}
\usepackage{amsmath}
\usepackage{braket}
\usepackage{subcaption,tikz}
\usetikzlibrary{arrows,decorations.markings,decorations.pathreplacing}

\makeatletter
\renewcommand\section{\@startsection {section}{1}{\z@}%
                                   {-3.5ex \@plus -1ex \@minus -.2ex}%nn
                                   {2.3ex \@plus.2ex}%
                                   {\normalfont\large\bfseries}}
\renewcommand\subsection{\@startsection{subsection}{2}{\z@}%
                                     {-3.25ex\@plus -1ex \@minus -.2ex}%
                                     {1.5ex \@plus .2ex}%
                                     {\normalfont\bfseries}}
\makeatother

\newcommand{\be}{\begin{equation}}
\newcommand{\ee}{\end{equation}}
\newcommand{\bea}{\begin{eqnarray}}
\newcommand{\eea}{\end{eqnarray}}

\newcommand{\al}{\alpha}
\renewcommand{\d}{\delta}
\newcommand{\e}{\epsilon}

\newcommand{\g}{\gamma}

\newcommand{\s}{\sigma}

\newcommand{\hlf}{\frac{1}{2}}

\newcommand{\non}{\nonumber}

\newcommand{\p}{\partial}

\newcommand{\R}{\mathbb{R}}
\newcommand{\rr}{\rightarrow}

\newcommand{\Z}{\mathbb{Z}}

\newcommand{\SL}{\operatorname{SL}}
\newcommand{\SO}{\operatorname{SO}}

\newcommand{\SU}{\operatorname{SU}}

\newcommand{\Tr}{\operatorname{Tr}}
\newcommand{\U}{\operatorname{U}}

\newcommand{\lp}{\left(}
\newcommand{\rp}{\right)}
\newcommand{\ls}{\left[}
\newcommand{\rs}{\right]}

\newcommand{\ol}[1]{{\overline{#1}}}
\newcommand{\ov}[1]{{\overline{#1}}}

\newcommand{\bc}{\bar{c}}
\newcommand{\bq}{\bar{q}}
\newcommand{\btau}{\bar{\tau}}
\newcommand{\bz}{\bar{z}}

\begin{document}

\renewcommand\thefootnote{\fnsymbol{footnote}}

\begin{titlepage}

\begin{center}

%\today
%\hfill         MIFPA-14-03\phantom{xxx}

\vskip 2 cm
{\Large \bf Orbifolds from Modular Orbits}\\
\vskip 1.25 cm {Daniel Robbins\footnote{email address: dgrobbins@albany.edu} and Thomas Vandermeulen\footnote{email address: tvandermeulen@albany.edu}}\\

{\vskip 0.5cm \it Department of Physics, University at Albany, \\ Albany, NY 12222, USA \\}

\end{center}
\vskip 2 cm

\begin{abstract}
\baselineskip=18pt
Given a two-dimensional conformal field theory with a global symmetry, we propose a method to implement an orbifold construction by taking orbits of the modular group.  For the case of cyclic symmetries we find that this approach always seems to be consistent, even in asymmetric orbifold cases where the usual construction does not yield a modular invariant theory; our approach keeps modular invariance manifest but may give a result that is equivalent to the original theory.  For the case that the symmetry is a subgroup of a continuous flavor symmetry, we can give explicit constructions of the spectrum, with twisted sectors corresponding to a non-standard group projection on an enlarged twisted sector Hilbert space.

\end{abstract}

\end{titlepage}

\pagestyle{plain}
%\baselineskip=18pt
% Try a wider skip
\baselineskip=19pt
%%%%%%%%%%%%%%%%%%%%%%%%%%%%%%%%%%%%%%%%%%%%%%%%%%%%%%%%%%%%%%%%%%%%%%%%%%%%%%

\addtocounter{footnote}{-2}
\renewcommand\thefootnote{\arabic{footnote}}

\section{Introduction}

Orbifolds have provided a large class of examples of two-dimensional conformal field theories.  In the original formulation~\cite{Dixon:1985jw,Dixon:1986jc}, orbifolds arose as the worldsheet theories describing strings that probe orbifold geometries - the result of quotienting a smooth manifold by an isometry subgroup with fixed points.  The most typical case starts with a free theory (a sigma model whose target space is a torus), and then constructs the quotient theory by simultaneously restricting to states that are invariant under the isometries, and by including new twisted sectors in which strings are closed only up to the quotient.  This procedure can be implemented at the level of the path integral, as we review in section \ref{sec:OrbPFs}.

In time, the term orbifold expanded in the physics literature~\cite{Narain:1986qm,Narain:1990mw,KlemmSchmidt} to refer to a broader class of constructions in which one quotients a conformal field theory by a global symmetry.  From this perspective, we can view an orbifold as simply gauging some (usually discrete) global symmetry of a CFT.  For the free theories mentioned above, we can look beyond the symmetries that correspond to isometries of the torus and consider more general quotients by symmetries that act differently on left- and right-movers~\cite{Narain:1986qm,Narain:1990mw,AokiDHokerPhong}.  These asymmetric orbifolds can be constructed using similar methods to the symmetric case, but the results are not always modular invariant.  Typically one discards the examples that fail to be modular invariant, declaring that the corresponding global symmetries simply can't be gauged.  In cases where there is no Lagrangian description of the parent theory, so that a path integral formulation is not available, then it is not always clear how to construct the orbifold theory.

An alternative route towards defining orbifolds in this more general context is to use modular invariance itself as a guiding principle.  In~\cite{AokiDHokerPhong}, such an approach is used to analyze asymmetric orbifolds of free boson parent theories, and an algorithm is presented which can be used in more general settings as well.  

There are a couple of ways that this procedure can go wrong.  If $H^2(G,\U(1))\ne 0$, then the full orbifold partition function is not in the orbit of the untwisted sector; in this case one must provide extra information (a choice of discrete torsion) to specify the orbifold theory, and, more seriously from our point of view, the disjoint orbits cannot be obtained from the parent theory and genus one modular invariance alone.  For cyclic groups, this is not an issue ($H^2(\Z_N,\U(1))=0$), and these will be our main focus in this paper.  A second pitfall can occur when $H^3(G,\U(1))\ne 0$.  In this case the symmetry $G$ could be anomalous.  Since orbifolding by a symmetry $G$ is essentially gauging that symmetry, then we would expect that an anomalous symmetry simply cannot be gauged.  In particular, we have $H^3(\Z_N,\U(1))\cong\Z_N$, so cyclic groups can be anomalous.

This paper then starts with a point of confusion.  It turns out that $H^2(\Z,\U(1))\cong H^3(\Z,\U(1))\cong 0$, so it should be that $\Z$-orbifolds are consistent (non-anomalous) and unambiguous (no discrete torsion).  So what prevents us from promoting a $\Z_N$ global symmetry (possibly anomalous) to a non-anomalous $\Z$ symmetry in which some elements act trivially, and then performing the orbifold.  Orbifolds in which the group acts ineffectively have been studied in~\cite{Pantev:2005rh,Hellerman:2006zs}.  In general the result may take the form of a direct sum of disconnected theories.  When discrete torsion is in play, the disconnected copies can involve different choices of discrete torsion.  For the cyclic groups we mostly consider, the disconnected copies will be identical.

The main objective of this paper is to explore this idea, of extending a $\Z_N$ symmetry to a non-anomalous $\Z$ symmetry to perform the orbifold.  This is very much in the spirit of~\cite{Witten:2016cio,Wang:2017loc}, which examines anomaly-free extensions of finite groups in the context of symmetry protected topological states.  We will find that for any cyclic symmetry we can construct a modular invariant orbifold partition function with all expected properties (non-negative integer coefficients, unique vacuum, etc.).  For cases that would have been anomalous, the resulting orbifold partition functions don't typically correspond to new theories, but rather reconstruct either the original parent theory or a different consistent orbifold thereof.  Another observation that emerges from our analysis is that even in non-anomalous orbifolds, one may need to use a non-standard projection in the enlarged twisted sector Hilbert spaces (i.e.\ states are not invariant under the group, but should transform in a particular projective representation).

We will mainly consider particular examples, and so we don't have general proofs that our proposal always results in consistent partition functions.  However, for the case that the quotient is by the subgroup of a continuous $\U(1)$ symmetry, we are able to use the technology of flavored partition functions to explicitly demonstrate many of the claimed properties, giving explicit expressions for all the partial traces and interpreting the twisted sector partition functions as projections on a set of enlarged Hilbert spaces.

The plan of the paper is as follows.  Section \ref{sec:OrbPFs} lays out our basic conventions for orbifold partition functions and partial traces.  Section \ref{sec:Main} contains our main proposal, in the form of a procedure for the construction of orbifold partition functions.  We follow this with a discussion of potential issues the formalism might face and some explicit examples of its implementation.  In section \ref{sec:Defects} we cast our proposal in the language of topological defect lines (TDLs), and in particular make more direct contact with the potential anomaly in our symmetry group.  Section \ref{sec:Flavored} focuses on theories with continuous symmetries, in which case we can introduce additional machinery in flavored partition functions.  This allows us to implement our procedure in an explicit form that applies to any such theory of this type.  The prototypical example of such a theory is the free boson with its U(1)$\times$U(1) symmetry, so in section \ref{sec:Boson} we examine these theories in some detail, showing how a variety of orbifold partition functions emerge as special cases of the calculations of section \ref{sec:Flavored}.  Section \ref{sec:TwoGenerator} demonstrates how our proposal can be iterated to encompass groups with multiple generators; we do not tackle this extension in full generality, but content ourselves with an example.  Section \ref{sec:Conclusion} recaps our conclusions and discusses some possible directions for future work.

\section{Orbifold partition functions and partial traces}
\label{sec:OrbPFs}

We begin with a consistent, unitary, two-dimensional conformal field theory which we will call the parent theory, that has some global symmetry that commutes with the Virasoro generators.  We assume that we know everything there is to know about the parent theory, %\footnote{We should emphasize that we are only considering local data in this paper.  We don't worry about any possible spectrum of extended objects, including boundaries.}, 
including the spectrum of primary operators, all correlation functions, etc.  Mostly we will restrict to theories with discrete spectra, though we will relax this in section \ref{subsec:ZOrbifold} in order to discuss the non-compact boson.  We would like to understand when there is a sensible ``orbifold procedure'' for such theories, i.e.\ a method which roughly corresponds to taking the quotient of the parent theory by some subgroup $G$ of the global symmetries to construct a new CFT called the orbifold theory.  We will be concerned only with quotients by discrete subgroups of the global symmetry group (and in fact we will later restrict to the case of cyclic subgroups).  The new theory will be called the orbifold theory.

As a first step, we would like our new theory to include all the states of the parent theory which are invariant under the chosen subgroup $G$.  At the level of the partition function, this means that our orbifold theory should include a sector, called the untwisted sector, of such invariant states, and the contribution of this sector to the orbifold partition function is given by inserting a projector onto invariant states in the partition function,
\be
Z_1=\Tr_{\mathcal{H}_P}\lp\Pi_Gq^{L_0-\tfrac{c}{24}}\bq^{\ol{L}_0-\tfrac{\bc}{24}}\rp,
\ee
where $\mathcal{H}_P$ is the Hilbert space of states in the parent theory.  In the case that $G$ is finite, the projector $\Pi_G$ can be written explicitly as
\be
\Pi_G=\frac{1}{|G|}\sum_{g\in G}\rho(g),
\ee
where $\rho:G\rr\U(\mathcal{H}_P)$ is the representation of the group $G$ on $\mathcal{H}_P$.  If we define {\it{untwisted sector partial traces}},
\be
Z_{1,g}:=\Tr_{\mathcal{H}_P}\lp\rho(g)q^{L_0-\tfrac{c}{24}}\bq^{\ol{L}_0-\tfrac{\bc}{24}}\rp,
\ee
then the quantity $Z_1$, which is called the {\it{untwisted sector partition function}} of the orbifold can be written as
\be
Z_1=\frac{1}{|G|}\sum_{g\in G}Z_{1,g}.
\ee

Unfortunately, $Z_1$ is not generally modular invariant.  There's no problem with the $\tau\rr\tau+1$ transformation; the partition function will be invariant under this if and only if (at least in cases where the CFT has a discrete spectrum) every state in the theory has integral spin, $h-\ol{h}\in\Z$.  Since the parent theory was modular invariant by assumption, and $Z_1$ simply traces over a subset of the original states, all the untwisted sector states will satisfy this condition.  On the other hand, we do not expect that $Z_1(-1/\tau,-1/\btau)=Z_1(\tau,\btau)$.  

The problem, and a way to fix it, is easiest to understand when one has access to a path integral formulation, for instance when the parent theory is a theory of free fields $\varphi$ with action $S[\varphi]$.  We review this story in appendix \ref{app:PIFormulation}.  The main takeaway is that one can define {\it{partial traces}} for any pair of commuting elements $g$ and $h$,
\be
Z_{h,g}(\tau,\btau)=\Tr_{\mathcal{H}_h}\ls\rho_h(g)q^{L_0-\tfrac{c}{24}}\bq^{\bar{L}_0-\tfrac{\bar{c}}{24}}\rs,
\ee
where $\mathcal{H}_h$ is the {\it{Hilbert space of states twisted by $h$}} and $\rho_h$ is the representation of the group $G$ acting on this Hilbert space.  Then we can build the {\it{twisted sector partition function}}, which when $G$ is finite and abelian is given by
\be
Z_h(\tau,\btau)=\frac{1}{|G|}\sum_{g\in G}Z_{h,g}(\tau,\btau),
\ee
corresponding to $G$-invariant states in the $h$-twisted sector.

The partial traces should transform according to equation (\ref{Appeq:ModTransOrbTraces}),
\be
\label{eq:ModTransOrbTraces}
Z_{h,g}(\frac{a\tau+b}{c\tau+d},\frac{a\btau+b}{c\btau+d})=Z_{g^{-c}h^a,g^dh^{-b}}(\tau,\btau),\qquad\mathrm{for}\qquad\lp\begin{matrix}a & b \\ c & d\end{matrix}\rp\in\SL(2,\Z).
\ee
And this means that an orbifold partition function built as
\be
Z_G=\sum_{h\in G}Z_h=\frac{1}{|G|}\sum_{h,g\in G}Z_{h,g},
\ee
is modular invariant (where again we have specialized to the finite abelian case).

When we don't have access to a path integral formulation of our theory, there does not seem to be a direct method to construct the twisted sector Hilbert spaces or the full partition function.  However, we can use modular invariance as a guiding principle, and attempt to {\it{define}} twisted sector partial traces by the transformation rule (\ref{eq:ModTransOrbTraces}), when that is possible, as we will describe in the next section.

\section{Method of modular orbits}
\label{sec:Main}

\subsection{Review of Aoki, D'Hoker, and Phong}

In~\cite{AokiDHokerPhong}, Aoki, D'Hoker, and Phong propose an approach to the construction of orbifolds which puts modular invariance at the forefront.  To illustrate their proposal, we will restrict to finite abelian groups $G$.  Those authors state two principles as a starting point.  First, they assume there exist enlarged twisted sector Hilbert spaces $\mathcal{H}_h$ for each $h\in G$, and a representation $\rho_h:G\rightarrow\U(\mathcal{H}_h)$ of $G$ on that Hilbert space, so that we can define partial traces as we did in the path integral case,
\be
Z_{h,g}(\tau,\btau)=\Tr_{\mathcal{H}_h}\ls\rho_h(g)q^{L_0-\tfrac{c}{24}}\bq^{\bar{L}_0-\tfrac{\bar{c}}{24}}\rs.
\ee
Secondly, they assume that these partial traces transform covariantly under modular transformations,
\be
%\label{eq:ModTransOrbTracesRedux}
Z_{h,g}(\frac{a\tau+b}{c\tau+d},\frac{a\btau+b}{c\btau+d})=Z_{g^{-c}h^a,g^dh^{-b}}(\tau,\btau),\qquad\mathrm{for}\qquad\lp\begin{matrix}a & b \\ c & d\end{matrix}\rp\in\SL(2,\Z).
\ee
For consistency, they insist on level matching, i.e.\ that if $h$ has order $n_h$ in $G$, then
\be
L_0-\ov{L}_0\in\frac{1}{n_h}\Z,
\ee
for all states in the enlarged twisted sector Hilbert space $\mathcal{H}_h$.

With these assumptions, then an orbifold partition function
\be
Z_G=\frac{1}{|G|}\sum_{g,h\in G}Z_{h,g},
\ee
is necessarily modular invariant.  Note that the twisted sector partition functions again involve projections,
\be
Z_h=\frac{1}{|G|}\sum_{g\in G}Z_{h,g}=\Tr_{\mathcal{H}_h}\ls\Pi_Gq^{L_0-\tfrac{c}{24}}\bq^{\bar{L}_0-\tfrac{\bar{c}}{24}}\rs.
\ee

The authors of~\cite{AokiDHokerPhong} next outline an algorithm for constructing the orbifold theory, which we shall paraphrase,
\begin{enumerate}
\item Use the knowledge of the parent theory to construct the untwisted sector partial traces $Z_{1,g}$.
\item Use modular transformations to compute $Z_{h,1}(\tau,\btau)=Z_{1,h}(-1/\tau,-1/\btau)$.  From these, deduce the spectrum of the enlarged twisted sector Hilbert space $\mathcal{H}_h$.
\item Further modular transformations furnish $Z_{h,h^k}(\tau,\btau)=Z_{h,1}(\tau-k,\btau-k)$, which can also be used to deduce the action of $h^k$ on $\mathcal{H}_h$.
\item Deduce the other twisted sector partial traces $Z_{h,g}$ by extrapolating from the actions of $h^k$ on the $\mathcal{H}_h$ (for instance some of these can be obtained by further modular transformations).
\end{enumerate}
Note that it may not always be clear how to implement the final step.

\subsection{Our proposal}
\label{sec:proposal}

Our proposal is similar in spirit to that of~\cite{AokiDHokerPhong}, and agrees with it in cases that they declare to be consistent, but we differ in a couple of points.  In this paper we will mainly focus on cyclic groups $G=\langle g\rangle\cong\Z_N$, and we will use additive notation for the group elements, $g^n\sim n$.  Our algorithm can be summarized as follows,
\begin{enumerate}
\item Use the knowledge of the parent theory to construct the untwisted sector partial traces $Z_{0,n}$.
\item By treating the group as $\Z$, which acts by $g$ but with $g^{kN}$ acting trivially for all integers $k$, we can apply modular transformations to the untwisted sector partial traces and obtain all partial traces $Z_{m,n}$.  The subscripts may not be periodic modulo $N$ (but will be periodic modulo $N^2$).
\item Construct the twisted sector partition functions
\be
Z_m(\tau,\btau)=\frac{1}{N^2}\sum_{n=0}^{N^2-1}Z_{m,n}(\tau,\btau).
\ee
Here $m$ will be periodic modulo $KN$ for some integer $1\le K\le N$, and we can construct the full orbifold partition function as
\be
\label{eq:SectorSumK}
Z_G(\tau,\btau)=\sum_{m=0}^{KN-1}Z_m(\tau,\btau).
\ee
Here $K$ will be the order of the anomaly of $G$, as an element of $H^3(G,\U(1))\cong\Z_N$ (explicitly, if $\al$ is the anomaly, then\footnote{As usual, $\gcd(m,n)$ represents the largest positive integer which divides the integers $m$ and $n$.  In particular, $\gcd(0,n)=n$.} $K=N/\gcd(N,\al)$).
%(Taking $K=N$ will give us a nice, modular invariant partition function, but may lead to multiple copies of the vacuum state.)
\end{enumerate}

There are of course some key differences between our approach and the one of ~\cite{AokiDHokerPhong}.  First of all, we remain agnostic on the subject of whether the $Z_{m,n}(\tau,\btau)$ necessarily have an interpretation as the trace of $\rho_h(n)$ in some enlarged twisted Hilbert space $\mathcal{H}_m$.  We do of course still expect that the full twisted sector partition function $Z_m(\tau,\btau)$ does have an interpretation as a trace over a some Hilbert space $\widehat{\mathcal{H}}_m$, but it may or may not be the case that $\widehat{\mathcal{H}}_m=\Pi_G(\mathcal{H}_m)$ for some $\mathcal{H}_m$ and $G$-action.

Secondly, we do not need to impose level-matching or equivalent constraints in order to get a consistent procedure.  By construction, our procedure always gives a modular invariant object which we would like to interpret as the orbifold partition function (in fact it gives a collection of twisted sector partition functions whose sum is the full partition function).  Although we have not actually constructed the twisted sector Hilbert spaces, much less given consistent OPEs among twisted sector operators, in every example we look at, the twisted sector partition functions are consistent with a Hilbert space interpretation (i.e.\ all the coefficients in the $q$ expansion are positive integers), and in the case where the orbifold group lies inside a continuous global symmetry, we can prove this property.

For cases that would fail the level-matching condition, i.e.\ cases in which the symmetry group $G$ has an anomaly, our construction seems to give something consistent, but not new.  Frequently one finds that the orbifold partition function is equal to the original parent theory partition function, but organized in a different way.  In other cases, our construction gives something equivalent to a different orbifold of the parent theory which is consistent in the traditional sense.

\subsection{Potential pitfalls to the procedure}
\label{sec:pitfalls}

To discuss the obstacles that may face us, and to work towards making our proposal more explicit, let us again return to the case of a general group $G$.  For a given parent theory, we can always compute all the $Z_{1,g}(\tau,\btau)$, by simply inserting $\rho(g)$ and tracing over $\mathcal{H}_P$.  Our proposal is to treat the transformation rule (\ref{eq:ModTransOrbTraces}) as fundamental and to build the full partition function by taking $\SL(2,\Z)$ orbits of the untwisted sector partition traces.  There are several ways that this procedure could potentially run into trouble:

\begin{itemize}
\item[(i)]%\label{it:DT} 
It might be that some of the required partial traces $Z_{h,g}$ do not lie in the $\SL(2,\Z)$ orbit of any of the untwisted partial traces, so we do not generate all possible partial traces by taking modular transformations.  Summing over the orbits of the untwisted partial traces will still generate a modular invariant candidate partition function, but it is well known~\cite{VafaTorsion} that higher genus considerations force us to include the missing orbits, possibly with some arbitrary choices of phase known as discrete torsion.  This discrete torsion is classified by the group cohomology $H^2(G;\U(1))$ (where we take the trivial group action on $\U(1)$).
\item[(ii)]%\label{it:LM} 
There may be inconsistencies in the identification of partial traces, at least if we assume that they are labeled by elements of the original symmetry group $G$.  This is actually related to a potential anomaly in the group $G$, which is classified by $H^3(G,\U(1))$.
%We will discuss this issue in more detail below.
\item[(iii)]%\label{it:HS} 
Even if the issues above are resolved or absent, the modular invariant quantity that we construct might not admit an interpretation as a partition function, i.e.\ there may be no Hilbert space of states over which we could trace $q^{L_0-\tfrac{c}{24}}\bq^{\bar{L}_0-\tfrac{\bar{c}}{24}}$ and get the given result.  This would be most obvious if the coefficients in a $q$ expansion are not positive integers, but could be more problematic even than this.  The only truly convincing construction of the orbifold would require a complete specification of the spectrum of operators and their OPEs.
\end{itemize}

The first item above is well-understood.  Consider the group $G=\Z_2\times\Z_2$, for which $H^2(\Z_2\times\Z_2;\U(1))\cong\Z_2$ (in general $H^2(\Z_n\times\Z_m;\U(1))\cong\Z_{\gcd(n,m)}$).  Then under modular transformations the partial traces arrange into five disjoint orbits,
\be
\begin{matrix}
\{Z_{(0,0),(0,0)}\},\\ \{Z_{(0,0),(0,1)},Z_{(0,1),(0,0)},Z_{(0,1),(0,1)}\},\\ \{Z_{(0,0),(1,0)},Z_{(1,0),(0,0)},Z_{(1,0),(1,0)}\},\\ \{Z_{(0,0),(1,1)},Z_{(1,1),(0,0)},Z_{(1,1),(1,1)}\},\\ \{Z_{(0,1),(1,0)},Z_{(0,1),(1,1)},Z_{(1,0),(0,1)},Z_{(1,0),(1,1)},Z_{(1,1),(0,1)},Z_{(1,1),(1,0)}\},\end{matrix}
\ee
where we have used additive notation for the group elements.  We see that the untwisted sector partial traces (i.e.\ the ones with $h=(0,0)$) only give us the first four orbits, and none of the partial traces in the final orbit can be obtained by one-loop modular transformations from the untwisted sector partial traces.  We will discuss how to apply our proposal to groups $\Z_N\times\Z_{N'}$ in section \ref{sec:TwoGenerator}.

On the other hand, for all cyclic groups, $G\cong\Z_n$ or $G\cong\Z$, we have $H^2(G;\U(1))\cong 0$, and all partial traces are covered by the orbits of the untwisted sector.  For these groups, the proposed method of orbits should be enough to construct all partial traces.  Most of this paper will focus on the case of cyclic groups, in which case item (i) will not be a concern.

Item (ii) is also understood in various circumstances.  Here the issue is that, because of relations among group elements, the determination of partial traces may be ambiguous, and hence inconsistent.  Recall the transformation rule (\ref{eq:ModTransOrbTraces}),
\be
\label{eq:ModTransOrbTracesRedux}
Z_{h,g}(\frac{a\tau+b}{c\tau+d},\frac{a\btau+b}{c\btau+d})=Z_{g^{-c}h^a,g^dh^{-b}}(\tau,\btau),\qquad\mathrm{for}\qquad\lp\begin{matrix}a & b \\ c & d\end{matrix}\rp\in\SL(2,\Z).
\ee
Clearly $Z_{1,1}(\tau,\btau)$ should be modular invariant, but we already have this condition since it is simply the partition function of the original parent theory, which was modular invariant by assumption.  What about the other untwisted sector partial traces, $Z_{1,g}(\tau,\btau)$?  Since the symmetry $g$ must commute with the Virasoro algebra, we can simultaneously diagonalize $\rho(g)$, $L_0$, and $\bar{L}_0$, and write
\be
\label{eq:weights}
Z_{1,g}(\tau,\btau)=q^{-\tfrac{c}{24}}\bq^{-\tfrac{\bar{c}}{24}}\sum_{\mathrm{states}\ i}g_iq^{h_i}\bq^{\bar{h}_i},
\ee
where $g_i$ are the eigenvalues of $\rho(g)$, and $h_i$ and $\bar{h}_i$ are the left and right conformal weights.  Invariance of the original partition function under the transformation $\tau\rr\tau+1$ enforces the conditions $c-\bar{c}\in 24\Z$ and $h_i-\bar{h}_i\in\Z$ for all states.  But these conditions are also sufficient to ensure that $Z_{1,g}(\tau,\btau)$ is invariant under $\tau\rr\tau+1$, as can be seen from the expression (\ref{eq:weights}) above.  In the case $G\cong\Z$, this is the only invariance we expect from (\ref{eq:ModTransOrbTracesRedux}), since setting $g^{-c}h^a=h$, $g^dh^{-b}=g$ for $h=1$ enforces $c=0$, and the corresponding $\SL(2,\Z)$ transformation will be $\tau\rr\tau+b$.  In general, if there are further relations on the generators of our group, for instance if $g^n=1$ for some $n>0$, then consistency requires more.  

Consider the case $G\cong\Z_2=\{0,1\}$.  Suppose we can compute $Z_{0,0}(\tau,\btau)$ and $Z_{0,1}(\tau,\btau)$.  $Z_{0,0}(\tau,\btau)$ is modular invariant and $Z_{0,1}(\tau,\btau)$ is invariant under $\tau\rr\tau+1$.  Then we might also expect to have
\be
Z_{0,1}(\frac{a\tau+b}{c\tau+d},\frac{a\btau+b}{c\btau+d})=Z_{-c,d}(\tau,\btau),
\ee
where $c$ and $d$ in the subscript on the right are taken modulo $2$.  Also note that the condition $ad-bc=1$ means that $c$ and $d$ are not both even.  By the usual expectations of orbifold constructions, we would find an inconsistency unless $Z_{0,1}(\tau,\btau)$ is invariant under the subgroup of modular transformations with $c$ even.  This subgroup is generated by the elements\footnote{To actually generate the subgroup of $\SL(2,\Z)$ as opposed to $\operatorname{PSL}(2,\Z)$, we would also need to include $\lp\begin{smallmatrix}-1 & 0 \\ 0 & -1\end{smallmatrix}\rp$ as a generator.} $\lp\begin{smallmatrix}1 & 1 \\ 0 & 1\end{smallmatrix}\rp$ and $\lp\begin{smallmatrix}1 & 0 \\ 2 & 1\end{smallmatrix}\rp$.  The first of these is simply the $\tau\rr\tau+1$ transformation and, as noted, invariance under that is guaranteed.  The second condition is non-trivial.  Thus we would conclude, according to the usual rules of orbifold construction, that the orbifold was inconsistent unless
\be
\label{eq:Z2MICond}
Z_{0,1}(\frac{\tau}{2\tau+1},\frac{\btau}{2\btau+1})=Z_{0,1}(\tau,\btau).
\ee
If this condition does hold, then we can define
\bea
\label{eq:Z2TwistedSectorPFDefs}
Z_{1,0}(\tau,\btau) &=& Z_{0,1}(-1/\tau,-1/\btau),\non\\
Z_{1,1}(\tau,\btau) &=& Z_{1,0}(\tau+1,\btau+1)=Z_{0,1}(-1/(\tau+1),-1/(\btau+1)),
\eea
and the manifestly modular invariant partition function
\bea
Z_{\Z_2}(\tau,\btau) &=& \hlf\lp Z_{0,0}(\tau,\btau)+Z_{0,1}(\tau,\btau)+Z_{1,0}(\tau,\btau)+Z_{1,1}(\tau,\btau)\rp\non\\
&=& Z_0(\tau,\btau)+Z_1(\tau,\btau),
\eea
where we have defined the untwisted and twisted sector partition functions,
\be
Z_0(\tau,\btau)=\hlf\lp Z_{0,0}(\tau,,\btau)+Z_{0,1}(\tau,\btau)\rp,\quad Z_1(\tau,\btau)=\hlf\lp Z_{1,0}(\tau,\btau)+Z_{1,1}(\tau,\btau)\rp.
\ee

Note that the condition (\ref{eq:Z2MICond}) can also be given a more familiar interpretation.  Since
\be
\label{eq:tauId}
\frac{\lp -\frac{1}{\tau+2}\rp}{2\lp -\frac{1}{\tau+2}\rp+1}=\frac{-1}{\tau},
\ee
we have
\be
\label{eq:TradLevelMatching}
Z_{1,0}(\tau+2,\btau+2)=Z_{0,1}(\frac{-1}{\tau+2},\frac{-1}{\btau+2})=Z_{0,1}(\frac{-1}{\tau},\frac{-1}{\btau})=Z_{1,0}(\tau,\btau),
\ee
where the middle equality uses (\ref{eq:tauId}) and (\ref{eq:Z2MICond}).  So (\ref{eq:Z2MICond}) implies that $Z_{1,0}$ is invariant under $\tau\rr\tau+2$.  Conversely, if $Z_{1,0}$ defined by (\ref{eq:Z2TwistedSectorPFDefs}) has this invariance, then
\be
Z_{0,1}(\frac{\tau}{2\tau+1},\frac{\btau}{2\btau+1})=Z_{1,0}(-\frac{1}{\tau}+2,-\frac{1}{\btau}+2)=Z_{1,0}(-\frac{1}{\tau},-\frac{1}{\btau})=Z_{0,1}(\tau,\btau).
\ee
We can similarly show that $Z_{1,1}(\tau+2,\btau+2)=Z_{1,1}(\tau,\btau)$ is equivalent.  What is the interpretation of this version of the invariance?  It's that the twisted sector, before imposing $\Z_2$ invariance, must obey that $2(h-\bar{h})\in\Z$ for all states (of course after we project onto invariant states we have the usual stronger condition that $h-\bar{h}\in\Z$).  This {\it{level-matching condition}} can be easier to use when the twisted sectors have been constructed geometrically rather than in some abstract formulation.  There are similar conditions for more complicated groups $G$.

As an example of where the condition (\ref{eq:Z2MICond}) fails to hold, consider the compact free boson at self-dual radius, which is $R=1$ in our conventions.  The partition function of the boson is (for detailed conventions, see section \ref{sec:examples} below)
\be
Z=\left|\eta\right|^{-2}\sum_{x,y\in\Z}q^{\frac{1}{4}\lp x+y\rp^2}\bq^{\frac{1}{4}\lp x-y\rp^2}.
\ee
This theory has a global symmetry group of $(\SU(2)\times\SU(2))/\Z_2$, where the $\Z_2$ is the element $(-1,-1)\in\SU(2)\times\SU(2)$.  The diagonal subgroup $\SU(2)/\Z_2\cong\SO(3)$ acts symmetrically on left- and right-moving degrees of freedom, and orbifolds by any discrete subgroup of this $\SO(3)$ (or any subgroup that is conjugate to one of these inside $(\SU(2)\times\SU(2))/\Z_2$) are known to be consistent.  On the other hand, we can consider symmetries which are not conjugate to something symmetric.  For $\Z_2$, there is only one such subgroup, which is generated by (the equivalence class of) the element $(1,-1)\in\SU(2)\times\SU(2)$.  This symmetry can be thought of as a chiral shift of order two\footnote{As explained in~\cite{AokiDHokerPhong,HarveyMoore}, this is also what one obtains by acting with T-duality twice.}, and the corresponding untwisted sector partial trace is
\be
Z_{0,1}=\left|\eta\right|^{-2}\sum_{x,y}e^{i\pi\lp x-y\rp}q^{\frac{1}{4}\lp x+y\rp^2}\bq^{\frac{1}{4}\lp x-y\rp^2}.
\ee
It is easy to verify~\cite{AokiDHokerPhong,HarveyMoore} that this does not satisfy (\ref{eq:Z2MICond}).  We will look at cases involving free bosons in much more detail below.

%The main point of our paper is to propose that a situation in which (\ref{eq:Z2MICond}) (or its analog for other groups) fails to hold should not be viewed as an inconsistency, at least for cyclic groups.  Rather, we should formally consider the indices on the partial traces as belonging to a larger group, use the method of modular orbits to construct modular partition functions, and then interpret the resulting spectrum.

Item (iii) on our list of concerns seems to be much more difficult to get a handle on.  As we will see in examples below, it does not seem to arise as an issue in the examples we construct, but in order to really prove that our procedure is well-defined, we would need to construct the full orbifold theory.  Even simply showing that the twisted sector partition functions have good $q$ expansions seems very difficult to prove in general, though in the case of orbifolds by flavor subgroups we will be able to demonstrate this explicitly.

\subsection{Orbifold by $\Z$}
\label{subsec:ZOrbifold}

For the case that our group $G$ is the integers, then the concerns in items (i) and (ii) do not arise, for the simple reason that $H^2(\Z,\U(1))\cong H^3(\Z,\U(1))\cong 0$.  In terms of practical consequences, it means that (using additive notation for our elements of $\Z$) we can give a unique unambiguous definition for all $Z_{m,n}$ in terms of the untwisted sector partial traces $Z_{0,r}$.  Indeed, let $r=\gcd(m,n)$.  Next use the Euclidean algorithm to find integers $a$ and $b$ that satisfy $an+bm=r$.  Then the matrix $\lp\begin{smallmatrix}a & b \\ -m/r & n/r\end{smallmatrix}\rp$ is in $\SL(2,\Z)$ and we can define
\be
\label{eq:ModTrans}
Z_{m,n}(\tau,\btau)=Z_{0,r}(r\frac{a\tau+b}{n-m\tau},r\frac{a\btau+b}{n-m\btau}).
\ee
There is of course redundancy in the choice of $a$ and $b$, but it is precisely accounted for by the $\tau\rr\tau+1$ invariance of the $Z_{0,n}(\tau,\btau)$.

Let's use this prescription to obtain the compact free boson at radius $R$ as a $\Z$-orbifold of the non-compact free boson.  This is a bit tricky, since the non-compact boson has a continuum of primary states $|p\rangle$ labeled by a real momentum $p\in\R$ with weights $h=\bar{h}=p^2/4$, and this in turn means that the partition function is divergent.  The divergence will be proportional to the (infinite) volume of space, and it can be argued~\cite{Polchinski:1998rq} that we should write something along the lines of
\be
Z_{nc}(\tau,\btau)=\mathrm{Vol}(\R)\left|\eta(\tau)\right|^{-2}\int_{-\infty}^\infty\frac{dp}{2\pi}\lp q\bq\rp^{\frac{p^2}{4}}=\frac{\mathrm{Vol}(\R)}{2\pi\sqrt{\tau_2}}\left|\eta(\tau)\right|^{-2}.
\ee

Now we would like to orbifold this theory by the symmetry group $G\cong\Z$ generated by a translation by $2\pi R$, which acts on primary states as
\be
g\cdot\left|p\right\rangle=e^{2\pi iRp}\left|p\right\rangle,
\ee
and which commutes with oscillators.  From this description it is straightforward to compute the untwisted sector partial traces,
\be
Z_{0,n}(\tau,\btau)=\mathrm{Vol}(\R)\left|\eta(\tau)\right|^{-2}\int_{-\infty}^\infty\frac{dp}{2\pi}\,e^{2\pi iRnp}\lp q\bq\rp^{\frac{p^2}{4}}=\frac{\mathrm{Vol}(\R)}{2\pi\sqrt{\tau_2}}|\eta(\tau)|^{-2}e^{-\frac{\pi R^2n^2}{\tau_2}}.
\ee

According to our formula (\ref{eq:ModTrans}), we can next define
\be
Z_{m,n}(\tau,\btau)=Z_{0,r}(r\frac{a\tau+b}{n-m\tau},r\frac{a\btau+b}{n-m\btau}),
\ee
where $r=\gcd(m,n)$ and $an+bm=r$.  Now the combination $|\eta|^{-2}/\sqrt{\tau_2}$ is precisely modular invariant, while $\tau_2$ transforms as
%, so it will be the same for all cases; we only have to evaluate the exponent.  How about $\tau_2$?  Computing, we have
\be
\operatorname{Im}\left\{r\frac{a\tau+b}{n-m\tau}\right\}=-\frac{ir}{2}\left\{\frac{a\tau+b}{n-m\tau}-\frac{a\btau+b}{n-m\btau}\right\}=-\frac{ir}{2}\frac{(an+bm)(\tau-\btau)}{|n-m\tau|^2}=\frac{r^2\tau_2}{|n-m\tau|^2}.
\ee
Hence it follows that
\be
Z_{m,n}(\tau,\btau)=Z_{0,r}(r\frac{a\tau+b}{n-m\tau},r\frac{a\btau+b}{n-m\btau})=\frac{\mathrm{Vol}(\R)}{2\pi\sqrt{\tau_2}}|\eta(\tau)|^{-2}e^{-\frac{\pi R^2|n-m\tau|^2}{\tau_2}}.
\ee

To compute the partition function in the $m$-twisted sector, we are instructed to compute
\begin{multline}
Z_m(\tau,\btau)=\frac{1}{|\Z|}\sum_{n\in\Z}Z_{m,n}(\tau,\btau)=\frac{\mathrm{Vol}(\R)}{2\pi|\Z|\sqrt{\tau_2}}|\eta(\tau)|^{-2}\sum_{n\in\Z}e^{-\frac{\pi R^2|n-m\tau|^2}{\tau_2}}\\
=\frac{\mathrm{Vol}(\R)}{2\pi R|\Z|}|\eta(\tau)|^{-2}\sum_{k\in\Z}e^{-\pi\tau_2\lp R^2m^2+\frac{k^2}{R^2}\rp+2\pi i\tau_1mk}=|\eta(\tau)|^{-2}\sum_{k\in\Z}e^{-\pi\tau_2\lp R^2m^2+\frac{k^2}{R^2}\rp+2\pi i\tau_1mk},
\end{multline}
where we have included yet another infinite constant, $|G|=|\Z|$, we have performed a Poisson resummation in going from the first line to the second, and in the final step we claim that two wrongs can make a right by identifying
\be
\frac{\mathrm{Vol}(\R)}{2\pi R|\Z|}=1,
\ee
i.e.\ that the full volume of $\R$ is obtained by the $|\Z|$ translates, each of length $2\pi R$.
%If we allow ourselves to take the ratio $\rho_0/(R|\Z|)$ equal to one\footnote{An argument in Polchinski Volume 1 indeed suggests that $\rho_0\sim\operatorname{Vol(\R)}/2\pi$, so if we squint our eyes and say that $\operatorname{Vol(\R)}\sim 2\pi R\cdot|\Z|$, this is in perfect agreement.}, then we recognize this as precisely the $m$-winding sector contribution to the compact free boson partition function at radius $R$.  
Finally, summing over $m$, we recover the full partition function,
\be
Z_\Z(\tau,\btau)=\frac{1}{|\Z|}\sum_{n,m\in\Z}Z_{m,n}(\tau,\btau)=Z_R(\tau,\btau),
\ee
where $Z_R(\tau,\btau)$ is the partition function for a compact free boson of radius $R$, which can also be computed directly,
\be
Z_R(\tau,\btau)=\left|\eta(\tau)\right|^{-2}\sum_{x,y\in\Z}q^{\frac{1}{4}\lp\frac{x}{R}+yR\rp^2}\bq^{\frac{1}{4}\lp\frac{x}{R}-yR\rp^2}.
\ee

\subsection{Fibered CFTs}
\label{subsec:FiberedCFTs}

With the caveats about infinite constants, which didn't slow us down much in the case of the non-compact boson, it seems that a modular orbit construction of a $\Z$-orbifold should always be consistent, at least in the sense of avoiding pitfalls (i) and (ii).  There are two ways that we might use this observation to extend to any cyclic orbifold group.  One is the algorithm we proposed above, and which will be our focus through most of this paper, but let us now mention another route.  Suppose we have any parent CFT $\mathcal{T}$ with a symmetry $g$ and that we would like to orbifold by the group generated by $g$.  If $g$ has order $N$, then this group is $\Z_N$ when acting on the parent theory.  In order to get a $\Z$-action instead, we can take the theory $\mathcal{T}$ and tensor it with a single non-compact boson to get a new theory,
\be
\mathcal{T}'=\mathcal{T}\otimes\mathcal{T}_{nc},
\ee
and take the symmetry operator $g'$ to act as $g$ in the factor $\mathcal{T}$ and as $e^{2\pi iRp}$ (translation by $ 2\pi R$) in the non-compact boson.  This operator generates a global $\Z$ symmetry of $\mathcal{T}'$, and we can try to build the orbifold theory $\mathcal{T'}/\Z$.

This procedure was proposed in~\cite{Hellerman:2006tx} and the resulting quotiented theories were called ``fibered CFTs''.  By taking modular orbits we can unambiguously build all of the partial traces and from there obtain the partition function $Z_{\mathcal{T}'/\Z}(\tau,\btau)$ of the orbifold theory.

The most interesting part of the construction was the dependence of the partition function on the parameter $R$, the shift in the noncompact direction.  If we take $R\rr\infty$, then all twisted sectors become very massive, their contribution to the partition function is suppressed, and we find that the partition function matches the decoupled product theory, $Z_{\mathcal{T}'/\Z}\cong Z_\mathcal{T}Z_{\mathcal{T}_{nc}}$, as we would expect.  On the other hand, suppose we take the limit $R\rr 0$.  In that case it was found that $Z_{\mathcal{T}'/\Z}\cong Z_{\mathcal{T}/\widehat{G}}Z_{\mathcal{T}_{nc}}$, where $\widehat{G}$ is the largest non-anomalous subgroup of $G\cong\Z_N$!  The subgroup $\widehat{G}\cong \Z_{N/k}$ will be generated by $g^k$ for some $k$ which divides $N$.

\subsection{Critical Ising model}

The critical Ising model has three Virasoro primaries in its spectrum: the identity $1$ with $h=\bar{h}=0$, a state $\e$ with $h=\bar{h}=1/2$, and a state $\s$ with $h=\bar{h}=1/16$.  The partition function is thus
\be
Z_{Ising}=\left|\chi_0(\tau)\right|^2+\left|\chi_{\hlf}(\tau)\right|^2+\left|\chi_{\frac{1}{16}}(\tau)\right|^2.
\ee

There is a $\Z_2$ global symmetry of this theory which flips the sign of $\s$, leaving $1$ and $\e$ alone.  Thus, our untwisted sector partial traces are
\bea
Z_{0,0} &=& \hlf\lp\left|\chi_0(\tau)\right|^2+\left|\chi_{\hlf}(\tau)\right|^2+\left|\chi_{\frac{1}{16}}(\tau)\right|^2\rp,\\
Z_{0,1} &=& \hlf\lp\left|\chi_0(\tau)\right|^2+\left|\chi_{\hlf}(\tau)\right|^2-\left|\chi_{\frac{1}{16}}(\tau)\right|^2\rp.
\eea

Under $\tau\rr -1/\tau$, the Ising model characters transform as
\be
\lp\begin{matrix}\chi_0 \\ \chi_{\hlf} \\ \chi_{\frac{1}{16}}\end{matrix}\rp\longrightarrow\lp\begin{matrix}\hlf & \hlf & \frac{1}{\sqrt{2}} \\ \hlf & \hlf & -\frac{1}{\sqrt{2}} \\ \frac{1}{\sqrt{2}} & -\frac{1}{\sqrt{2}} & 0\end{matrix}\rp\lp\begin{matrix}\chi_0 \\ \chi_{\hlf} \\ \chi_{\frac{1}{16}}\end{matrix}\rp,
\ee
while under $\tau\rr\tau+1$, we have
\be
\chi_0\rr e^{-\frac{\pi i}{24}}\chi_0,\quad\chi_{\hlf}\rr -e^{-\frac{\pi i}{24}}\chi_{\hlf},\quad\chi_{\frac{1}{16}}\rr e^{\frac{\pi i}{8}-\frac{\pi i}{24}}\chi_{\frac{1}{16}}.
\ee

Then applying our method of modular orbits gives us the twisted sector partial traces\footnote{Note that $Z_{1,1}(\tau,\btau)=Z_{0,1}(\frac{1}{1-\tau},\frac{1}{1-\btau})$, but we can rewrite the latter as $Z_{1,0}(\tau-1,\btau-1)$ using our result for $Z_{1,0}$.},
\bea
Z_{1,0}(\tau,\btau) &=& Z_{0,1}(-1/\tau,-1/\btau)\non\\
&=& \left|\hlf\chi_0+\hlf\chi_{\hlf}+\frac{1}{\sqrt{2}}\chi_{\frac{1}{16}}\right|^2+\left|\hlf\chi_0+\hlf\chi_{\hlf}-\frac{1}{\sqrt{2}}\chi_{\frac{1}{16}}\right|^2\non\\
&& \qquad -\left|\frac{1}{\sqrt{2}}\chi_0-\frac{1}{\sqrt{2}}\chi_{\hlf}\right|^2\non\\
&=& \chi_0\ov{\chi}_{\hlf}+\chi_{\hlf}\ov{\chi}_0+\left|\chi_{\frac{1}{16}}\right|^2,\\
Z_{1,1}(\tau,\btau) &=& Z_{1,0}(\tau-1,\btau-1)\non\\
&=& -\chi_0\ov{\chi}_{\hlf}-\chi_{\hlf}\ov{\chi}_0+\left|\chi_{\frac{1}{16}}\right|^2.
\eea
Furthermore, the usual procedure is consistent because (\ref{eq:TradLevelMatching}) is satisfied by $Z_{1,0}$.  In other words, this symmetry is anomaly-free.

We then have the untwisted and twisted sector partition functions,
\bea
Z_0 &=& \hlf\lp Z_{0,0}+Z_{0,1}\rp=\left|\chi_0\right|^2+\left|\chi_{\hlf}\right|^2,\\
Z_1 &=& \hlf\lp Z_{1,0}+Z_{1,1}\rp=\left|\chi_{\frac{1}{16}}\right|^2.
\eea
So the full orbifold partition function simply reproduces the original partition function!

\subsection{Reflection and shift orbifolds of the free boson}
\label{sec:examples}

Many of the examples that we examine in this paper involve one or several compact free bosons.  We will focus more closely on these theories in section~\ref{sec:Boson}, using the tool of flavored partition functions.  But here we will establish some conventions and look at some simple orbifolds.  

The spectrum of the compact free boson at radius $R$ (we set $\al'=1$ so that in our conventions the self-dual radius is $R=\sqrt{\al'}=1$) consists of lowest weight states with fixed momentum and winding, labeled by integers $x$ and $y$, $|x,y\rangle$, and ($\U(1)$ current algebra) descendants obtained from these states by acting on them with left- and right-moving raising operators $\al_{-m}$, $\widetilde{\al}_{-m}$.  If we let $N_m$ and $\widetilde{N}_m$ represent the number of $\al_{-m}$ and $\widetilde{\al}_m$ operators respectively, then a general state is
\be
\left|x,y;\{N_1,N_2,\cdots\},\{\widetilde{N}_1,\widetilde{N}_2,\cdots\}\right\rangle=C_{x,y;\vec{N},\vec{\widetilde{N}}}\prod_{m=1}^\infty\al_{-m}^{N_m}\widetilde{\al}_{-m}^{\widetilde{N}_m}\left|x,y\right\rangle,
\ee
where $C_{x,y;\vec{N},\vec{\widetilde{N}}}$ is a normalization constant whose precise form we won't need except to note that $C_{-x,-y;\vec{N},\vec{\widetilde{N}}}=C_{x,y;\vec{N},\vec{\widetilde{N}}}$.  The physical Hilbert space only includes states with a finite number of raising operators, i.e.\ with $\sum_{m=1}^\infty(N_m+\widetilde{N}_m)<\infty$.  The corresponding partition function is then
\be
Z_R(\tau,\btau)=\left|\eta(\tau)\right|^{-2}\sum_{x,y\in\Z}q^{\tfrac{1}{4}\lp\tfrac{x}{R}+yR\rp^2}\bq^{\tfrac{1}{4}\lp\tfrac{x}{R}-yR\rp^2}.
\ee

For $R\ne 1$, this theory has a $(\U(1)\times\U(1))\rtimes\Z_2$ global symmetry group.  The group's elements can be labeled by two periodic angles $0\le\al,\beta\le 2\pi$ and a discrete parameter $\chi=\pm 1$.  Then the group multiplication and inversion are given by
\be
\lp\al_1,\beta_1;\chi_1\rp\cdot\lp\al_2,\beta_2;\chi_2\rp=\lp\al_2+\al_1\chi_2,\beta_2+\beta_1\chi_2;\chi_1\chi_2\rp,\qquad\lp\al,\beta;\chi\rp^{-1}=\lp -\chi\al,-\chi\beta;\chi\rp.
\ee
The action on states is
\be
\lp\al,\beta;\chi\rp\cdot\left|x,y;\vec{N},\vec{\widetilde{N}}\right\rangle=\chi^{\sum_{m=1}^\infty\lp N_m+\widetilde{N}_m\rp}e^{i\al x+i\beta y}\left|\chi x,\chi y;\vec{N},\vec{\widetilde{N}}\right\rangle.
\ee
The $\chi=+1$ elements act as translations (and dual translations), while the $\chi=-1$ elements are reflections.

We can easily write an expression for the trace over the Hilbert space with a group element inserted,
\bea
\label{eq:1BosonRotation}
Z_{1,\lp\al,\beta;1\rp}(\tau,\btau) &=& \left|\eta(\tau)\right|^{-2}\sum_{x,y\in\Z}e^{i\al x+i\beta y}q^{\tfrac{1}{4}\lp\tfrac{x}{R}+yR\rp^2}\bq^{\tfrac{1}{4}\lp\tfrac{x}{R}-yR\rp^2},\\
Z_{1,\lp\al,\beta;-1\rp}(\tau,\btau) &=& \left|q\right|^{-\tfrac{1}{12}}\lp\prod_{n=1}^\infty\frac{1}{\left|1+q^n\right|^2}\rp=\left|\frac{2\eta(\tau)}{\theta_2(\tau)}\right|.
\eea
In the latter case, the trace has localized onto states with $x=y=0$.

Let us now look at orbifolds generated by elements of order two (switching to additive notation for the group elements).  If our generator is a reflection (and all reflections are conjugate to each other), then we have
\bea
Z_{0,0}(\tau,\btau) &=& \left|\eta(\tau)\right|^{-2}\sum_{x,y\in\Z}q^{\tfrac{1}{4}\lp\tfrac{x}{R}+yR\rp^2}\bq^{\tfrac{1}{4}\lp\tfrac{x}{R}-yR\rp^2},\\
Z_{0,1}(\tau,\btau) &=& \left|\frac{2\eta(\tau)}{\theta_2(\tau)}\right|,\\
Z_{1,0}(\tau,\btau) &=& Z_{0,1}(-1/\tau,-1/\btau)=\left|\frac{2\eta(\tau)}{\theta_4(\tau)}\right|,\\
Z_{1,1}(\tau,\btau) &=& Z_{1,0}(\tau-1,\btau-1)=\left|\frac{2\eta(\tau)}{\theta_3(\tau)}\right|.
\eea
Here, the traditional level-matching is satisfied, so there is no more to the story.  We have recovered the partition function for the standard boson reflection orbifold,
\be
Z^{S^1/\Z_2}_R(\tau,\btau)=\hlf\lp Z_{0,0}+Z_{0,1}\rp+\hlf\lp Z_{1,0}+Z_{1,1}\rp.
\ee

Among the translation elements, there are three possible elements of order two.  Let's look at each of them in turn.

\begin{itemize}
\item[(i)] $g=(\pi,0;1)$

From (\ref{eq:1BosonRotation}), we have
\be
Z_{0,1}(\tau,\btau)=\left|\eta(\tau)\right|^{-2}\sum_{x,y\in\Z}\lp -1\rp^xq^{\tfrac{1}{4}\lp\tfrac{x}{R}+yR\rp^2}\bq^{\tfrac{1}{4}\lp\tfrac{x}{R}-yR\rp^2}.
\ee
Our procedure then gives (after performing a pair of Poisson resummations)
\be
Z_{1,0}(\tau,\btau)=Z_{0,1}(-1/\tau,-1/\btau)=\left|\eta(\tau)\right|^{-2}\sum_{x,y\in\Z}q^{\tfrac{1}{4}\lp\frac{x}{R}+(y-\hlf)R\rp^2}\bq^{\tfrac{1}{4}\lp\frac{x}{R}-(y-\hlf)R\rp^2},
\ee
and
\be
Z_{1,1}(\tau,\btau)=Z_{1,0}(\tau-1,\btau-1)=\left|\eta(\tau)\right|^{-2}\sum_{x,y\in\Z}\lp -1\rp^xq^{\tfrac{1}{4}\lp\frac{x}{R}+(y-\hlf)R\rp^2}\bq^{\tfrac{1}{4}\lp\frac{x}{R}-(y-\hlf)R\rp^2}.
\ee

Combining these results, the untwisted and twisted sector partition functions become (standard level matching is fine in this case)
\bea
Z_0(\tau,\btau) &=& \left|\eta(\tau)\right|^{-2}\sum_{x,y\in\Z}\frac{1+\lp -1\rp^x}{2}q^{\tfrac{1}{4}\lp\tfrac{x}{R}+yR\rp^2}\bq^{\tfrac{1}{4}\lp\tfrac{x}{R}-yR\rp^2}\non\\
&=& \left|\eta(\tau)\right|^{-2}\sum_{k,y\in\Z}q^{\tfrac{1}{4}\lp\tfrac{k}{R/2}+yR\rp^2}\bq^{\tfrac{1}{4}\lp\tfrac{k}{R/2}-yR\rp^2},\\
Z_1(\tau,\btau) &=& \left|\eta(\tau)\right|^{-2}\sum_{x,y\in\Z}\frac{1+\lp -1\rp^x}{2}q^{\tfrac{1}{4}\lp\frac{x}{R}+(y-\hlf)R\rp^2}\bq^{\tfrac{1}{4}\lp\frac{x}{R}-(y-\hlf)R\rp^2}\\
&=& \left|\eta(\tau)\right|^{-2}\sum_{k,y\in\Z}q^{\tfrac{1}{4}\lp\frac{k}{R/2}+(y-\hlf)R\rp^2}\bq^{\tfrac{1}{4}\lp\frac{k}{R/2}-(y-\hlf)R\rp^2},
\eea
where in each sector we end up with a projection onto terms with $x$ even, and so in the second line we have rewritten things in terms of $k=x/2$.  The two sectors now look identical except that $y$ in the untwisted sector is replaced by $(y-\hlf)$ in the twisted sector, so when we build the full partition function we simply have a sum over all half integer values, and so we can write
\be
\label{z2coordshift}
Z(\tau,\btau)=Z_0(\tau,\btau)+Z_1(\tau,\btau)=\left|\eta(\tau)\right|^{-2}\sum_{k,\ell\in\Z}q^{\tfrac{1}{4}\lp\frac{k}{R/2}+\frac{\ell R}{2}\rp^2}\bq^{\tfrac{1}{4}\lp\frac{k}{R/2}-\frac{\ell R}{2}\rp^2},
\ee
which is simply the partition function at radius $R'=R/2$.  In other words, this orbifold is the standard one which identifies the circle under a half-period translation, and takes us from the theory at radius $R$ to the theory at $R/2$.  Of course there is no problem with implementing this construction completely using the usual procedure discussed in section~\ref{sec:OrbPFs}.  We have simply shown that it can also be recovered using our procedure.

\item[(ii)] $g=(0,\pi;1)$

This case proceeds almost identically, with the roles of momentum and winding reversed (it is the T-dual of the previous case).  The result is the partition function at radius $R'=2R$.

\item[(iii)] $g=(\pi,\pi;1)$

Finally, we turn to the most non-standard case.  From  (\ref{eq:1BosonRotation}),
\be
Z_{0,1}(\tau,\btau)=\left|\eta(\tau)\right|^{-2}\sum_{x,y\in\Z}\lp -1\rp^{x+y}q^{\tfrac{1}{4}\lp\tfrac{x}{R}+yR\rp^2}\bq^{\tfrac{1}{4}\lp\tfrac{x}{R}-yR\rp^2}.
\ee
The untwisted sector partition function simply involves a projection onto invariant states, i.e.\ states for which $x+y$ is even.

Again, we easily obtain, by performing Poisson resummations
\be
Z_{1,0}(\tau,\btau)=Z_{0,1}(-1/\tau,-1/\btau)=\left|\eta(\tau)\right|^{-2}\sum_{x,y\in\Z}q^{\tfrac{1}{4}\lp\frac{x-\hlf}{R}+(y-\hlf)R\rp^2}\bq^{\tfrac{1}{4}\lp\frac{x-\hlf}{R}-(y-\hlf)R\rp^2}.
\ee
This result does not obey level-matching; it is not invariant under $\tau\rr\tau+2$, though it is invariant under $\tau\rr\tau+4$.  We find the following partial traces in this sector,
\bea
Z_{1,1}(\tau,\btau) &=& Z_{1,0}(\tau-1,\btau-1)\non\\
&=& -i\left|\eta(\tau)\right|^{-2}\sum_{x,y\in\Z}\lp -1\rp^{x+y}q^{\tfrac{1}{4}\lp\frac{x-\hlf}{R}+(y-\hlf)R\rp^2}\bq^{\tfrac{1}{4}\lp\frac{x-\hlf}{R}-(y-\hlf)R\rp^2}\\
Z_{1,2}(\tau,\btau) &=& Z_{1,0}(\tau+2,\btau+2)=-Z_{1,0}(\tau,\btau),\\
Z_{1,3}(\tau,\btau) &=& Z_{1,0}(\tau+1,\btau+1)=-Z_{1,1}(\tau,\btau).
\eea
If we write out the full twisted sector partition function, we find that everything cancels,
\be
Z_1(\tau,\btau)=\frac{1}{4}\lp Z_{1,0}(\tau,\btau)+Z_{1,1}(\tau,\btau)+Z_{1,2}(\tau,\btau)+Z_{1,3}(\tau,\btau)\rp=0.
\ee
This sector is empty!

How about the $g^2$-twisted sector?  Although $g^2$ acts trivially in the parent theory, we will find that this sector is not identical to the untwisted sector.  Our proposal gives
\bea
Z_{2,0}(\tau,\btau) &=& Z_{0,2}(-1/\tau,-1/\btau)=Z_{0,0}(\tau,\btau),\\
Z_{2,1}(\tau,\btau) &=& Z_{1,-2}(-1/\tau,-1/\btau)=-Z_{0,1}(\tau,\btau),\\
Z_{2,2}(\tau,\btau) &=& Z_{2,0}(\tau-1,\btau-1)=Z_{0,0}(\tau,\btau),\\
Z_{2,3}(\tau,\btau) &=& Z_{1,2}(-1/\tau,-1/\btau)=-Z_{0,1}(\tau,\btau).
\eea
Adding things together in this sector, we find
\begin{multline}
Z_2(\tau,\btau)=\frac{1}{4}\lp Z_{2,0}(\tau,\btau)+Z_{2,1}(\tau,\btau)+Z_{2,2}(\tau,\btau)+Z_{2,3}(\tau,\btau)\rp\\
=\hlf\lp Z_{0,0}(\tau,\btau)-Z_{0,1}(\tau,\btau)\rp.
\end{multline}
This sector contains the states of the parent theory with a projection onto $x+y$ odd!  Combined with the untwisted sector, which projected onto states with $x+y$ even, these two sectors contain precisely the states of the original parent theory.

The $g^3$-twisted sector works out in the same fashion as the $g$-twisted sector, and the projection kills everything.  And the $g^4$-twisted sector becomes again identical to the untwisted sector.  This is an example for which the symmetry is anomalous, represented by the nontrivial element in $H^3(\Z_2,\U(1))\cong\Z_2$~\cite{AokiDHokerPhong,HarveyMoore,Lin:2019kpn}.  This element of course has order two, and so the prescription of (\ref{eq:SectorSumK}) says that we should pick $K=2$ and put
\be
Z_G=\sum_{m=0}^3Z_m=Z_0+Z_2=Z_R,
\ee
and we find we have recovered the partition function for the original parent theory.

\end{itemize}

\subsection{Cyclic permutation orbifold}

Let's give one more example of a non-anomalous $\Z_2$ symmetry, to further illustrate our procedure.  Consider any CFT $\mathcal{T}$, and construct the tensor product of it with itself, $\mathcal{T}\otimes\mathcal{T}$.  The Hilbert space of states is simply the tensor product of two copies of the $\mathcal{T}$ Hilbert space, $\mathcal{H}=\mathcal{H}_\mathcal{T}\otimes\mathcal{H}_\mathcal{T}$, and there is a simple $\Z_2$ exchange symmetry which sends $|\psi_1\rangle\otimes|\psi_2\rangle$ to $|\psi_2\rangle\otimes|\psi_1\rangle$.  We can easily construct the untwisted sector partial traces,
\be
Z_{0,0}(\tau,\btau)=Z_\mathcal{T}(\tau,\btau)^2,\qquad Z_{0,1}(\tau,\btau)=Z_\mathcal{T}(2\tau,2\btau).
\ee
Here $Z_{0,1}$ is computed by observing that the only states that survive the trace with the exchange symmetry inserted are the states of the form $|\psi\rangle\otimes|\psi\rangle$, and these contribute a total weight of twice the weight of $|\psi\rangle$.

We can get the twisted sector partial traces by taking modular orbits,
\be
Z_{1,0}(\tau,\btau)=Z_{0,1}(-1/\tau,-1/\btau)=Z_\mathcal{T}(-2/\tau,-2/\btau)=Z_\mathcal{T}(\tau/2,\btau/2),
\ee
where in the final equality we used the modular invariance of $Z_\mathcal{T}$.  This $Z_{1,0}$ obeys level-matching, $Z_{1,0}(\tau+2,\btau+2)=Z_{1,0}(\tau,\btau)$, showing that the anomaly is vanishing.  We can then compute
\be
Z_{1,1}(\tau,\btau)=Z_{1,0}(\tau+1,\btau+1)=Z_\mathcal{T}(\tfrac{\tau+1}{2},\tfrac{\btau+1}{2}).
\ee

In total then, we find that the untwisted sector is given by the exchange invariant states, i.e.\ 
\be
Z_0(\tau,\btau)=\hlf Z_\mathcal{T}(\tau,\btau)^2+\hlf Z_\mathcal{T}(2\tau,2\btau),
\ee
and the twisted sector is
\be
Z_1(\tau,\btau)=\hlf\ls Z_\mathcal{T}(\tfrac{\tau}{2},\tfrac{\btau}{2})+Z_\mathcal{T}(\tfrac{\tau+1}{2},\tfrac{\btau+1}{2})\rs.
\ee
These results match those found, for instance, in \cite{Borisov:1997nc}.

\section{Partial traces from insertion of topological defects}
\label{sec:Defects}

A more modern version of our story is to associate each element $g$ of a symmetry group $G$ with a topological defect line (TDL) $\mathcal{L}_g$ (we will largely use the notation of~\cite{Chang:2018iay}).  These defects can join at junctions and can fuse together, with the fusion obeying the group multiplication law.  Laying down a network of these defect lines and junctions is equivalent to coupling the theory to a background gauge field for the symmetry group $G$.  In order for a physical result to only depend on the choice of a flat background connection, different networks with the same topology should give the same result; in particular the two different ways of resolving a degree-four junction into a pair of trivalent junctions should agree.  These different ways are classified by an 't Hooft anomaly measured by a class in the group cohomology $H^3(G,\U(1))$.  Then gauging, or orbifolding, our theory by $G$ amounts to summing over all possible background gauge field configurations, i.e.\ all possible networks of defects.  One usually declares that this is only a sensible procedure when the group is anomaly free.

So again the question arises of how this works if we find an anomaly free extension of our symmetry group.  In particular, we will again take a $\Z_N$ symmetry and then promote it to a $\Z$ symmetry.  We will propose a sense in which the $\Z$ symmetry can be sensibly gauged.

\subsection{Modular transformations}

\begin{figure}
\begin{subfigure}{0.5\textwidth}
\centering
\begin{tikzpicture}[scale=1.5]
   \draw[color=lightgray,->] (-0.5,0) -- (2.5,0);
   \draw[color=lightgray,->] (0,-0.5) -- (0,1.5);
   \draw[color=lightgray] (2,-0.5) -- (2,1.5);
   \draw[color=lightgray] (-0.5,1) -- (2.5,1);
   \draw (0,0) -- (2,0) -- (2,1) -- (0,1) -- (0,0);
   \draw[very thick,->] (0,0.5) -- (1,0.5);
   \draw[very thick] (1,0.5) -- (2,0.5);
   \draw[very thick,color=lightgray] (-0.5,0.5) -- (0,0.5);
   \draw[very thick,color=lightgray] (2,0.5) -- (2.5,0.5);
   \node at (-0.1,1.1) {$\tau$};
   \node at (1,0.7) {$g$};
\end{tikzpicture}
\caption{}
\end{subfigure}
\begin{subfigure}{0.5\textwidth}
\centering
\begin{tikzpicture}[scale=1.5]
   \draw[color=lightgray,->] (-0.5,0) -- (1.5,0);
   \draw[color=lightgray,->] (0,-0.5) -- (0,2.5);
   \draw[color=lightgray] (1,-0.5) -- (1,2.5);
   \draw[color=lightgray] (-0.5,2) -- (1.5,2);
   \draw (0,0) -- (1,0) -- (1,2) -- (0,2) -- (0,0);
   \draw[very thick,->] (0.5,0) -- (0.5,1);
   \draw[very thick] (0.5,1) -- (0.5,2);
   \draw[very thick,color=lightgray] (0.5,-0.5) -- (0.5,0);
   \draw[very thick,color=lightgray] (0.5,2) -- (0.5,2.5);
   \node at (-0.1,2.1) {$\tau'$};
   \node at (0.3,1) {$g$};
\end{tikzpicture}
\caption{}
\end{subfigure}
\caption{The untwisted sector partial traces $Z_{1,g}$ are given by inserting a group element in the trace.  If we think of the action of the group element as being implemented by a topological defect wrapping the spatial circle at a fixed time, then Figure (a) represents $Z_{1,g}$.  To get $Z_{g,1}$, we perform a $\tau\rr\tau'=-1/\tau$ modular transformation, resulting in a defect wrapping the time circle, as in Figure (b).}
\label{fig:TDLSTransformation}
\end{figure}
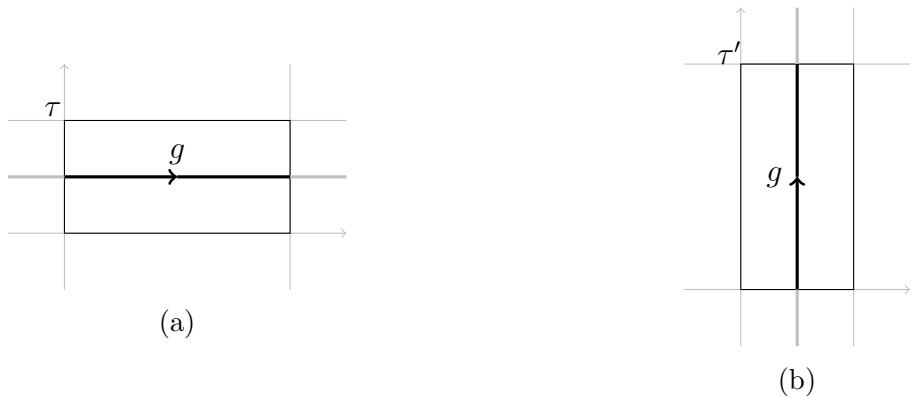

First we need to understand how our modular orbits prescription works in the language of TDLs.  To start with, a TDL labeled with a symmetry $g$ that stretches across the spatial cycle of a torus represents an insertion of (the representative of) $g$ into the trace, i.e.\ the left of Figure \ref{fig:TDLSTransformation} represents the untwisted sector partial trace $Z_{1,g}(\tau,\btau)$.  A $\tau\rr -1/\tau$ modular transformation swaps the spatial and temporal cycles, resulting in the right of Figure \ref{fig:TDLSTransformation}, representing the twisted sector partial trace $Z_{g,1}(\tau,\btau)$.

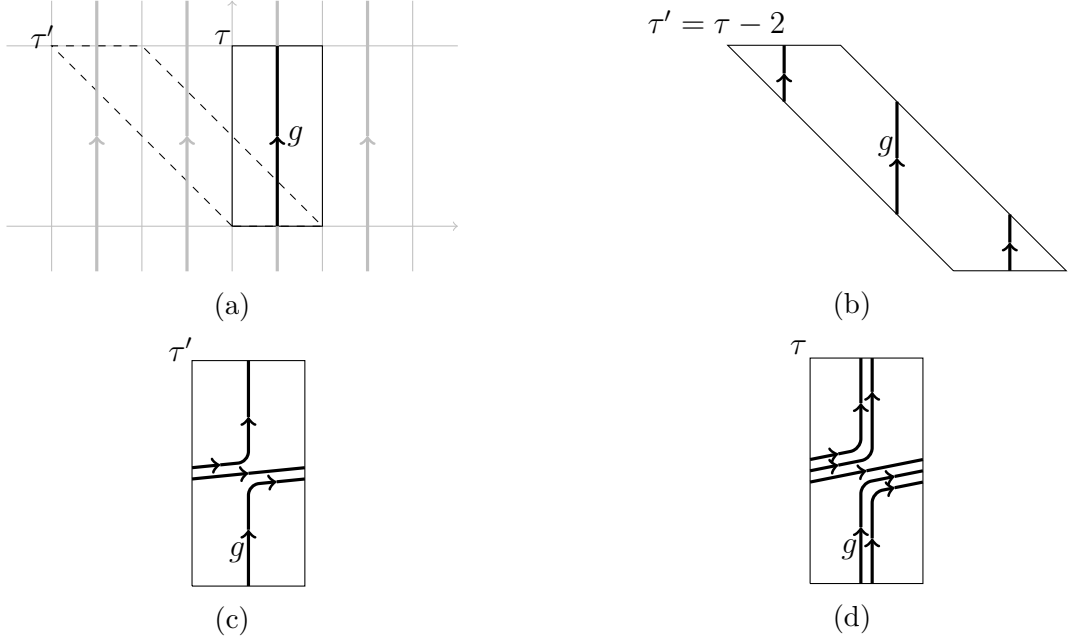
\begin{figure}[h]
\begin{subfigure}{0.5\textwidth}
\centering
\begin{tikzpicture}[scale=1.2]
   \draw[color=lightgray,->] (-2.5,0) -- (2.5,0);
   \draw[color=lightgray,->] (0,-0.5) -- (0,2.5);
   \draw[color=lightgray] (-2.5,2) -- (2.5,2);
   \draw[color=lightgray] (-2,-0.5) -- (-2,2.5);
   \draw[color=lightgray] (-1,-0.5) -- (-1,2.5);
   \draw[color=lightgray] (1,-0.5) -- (1,2.5);
   \draw[color=lightgray] (2,-0.5) -- (2,2.5);
   \draw (0,0) -- (1,0) -- (1,2) -- (0,2) -- (0,0);
   \node at (-0.1,2.1) {$\tau$};
   \draw[dashed] (0,0) -- (-2,2) -- (-1,2) -- (1,0) -- (0,0);
   \node at (-2.1,2.1) {$\tau'$};
   \draw[very thick,->] (0.5,0) -- (0.5,1);
   \draw[very thick] (0.5,1) -- (0.5,2);
   \draw[very thick,color=lightgray] (0.5,-0.5) -- (0.5,0);
   \draw[very thick,color=lightgray] (0.5,2) -- (0.5,2.5);
   \draw[very thick,color=lightgray,->] (-0.5,-0.5) -- (-0.5,1);
   \draw[very thick,color=lightgray] (-0.5,1) -- (-0.5,2.5);
   \draw[very thick,color=lightgray,->] (-1.5,-0.5) -- (-1.5,1);
   \draw[very thick,color=lightgray] (-1.5,1) -- (-1.5,2.5);
   \draw[very thick,color=lightgray,->] (1.5,-0.5) -- (1.5,1);
   \draw[very thick,color=lightgray] (1.5,1) -- (1.5,2.5);
   \node at (0.7,1) {$g$};
\end{tikzpicture}
\caption{}
\end{subfigure}
\begin{subfigure}{0.5\textwidth}
\centering
\begin{tikzpicture}[scale=1.5]
   \draw (0,0) -- (-2,2) -- (-1,2) -- (1,0) -- (0,0);
   \node at (-2.1,2.2) {$\tau'=\tau-2$};
   \draw[very thick,->] (0.5,0) -- (0.5,0.25);
   \draw[very thick] (0.5,0.25) -- (0.5,0.5);
   \draw[very thick,->] (-0.5,0.5) -- (-0.5,1);
   \draw[very thick] (-0.5,1) -- (-0.5,1.5);
   \draw[very thick,->] (-1.5,1.5) -- (-1.5,1.75);
   \draw[very thick] (-1.5,1.75) -- (-1.5,2);
   \node at (-0.6,1.1) {$g$};
\end{tikzpicture}
\caption{}
\end{subfigure}
\begin{subfigure}{0.5\textwidth}
\centering
\begin{tikzpicture}[scale=1.5]
   \draw (0,0) -- (1,0) -- (1,2) -- (0,2) -- (0,0);
   \node at (-0.1,2.1) {$\tau'$};
   \draw[very thick,->] (0.5,0) -- (0.5,0.5);
   \draw[very thick,rounded corners,->] (0.5,0.5) -- (0.5,0.9) -- (0.75,0.925);
   \draw[very thick] (0.75,0.925) -- (1,0.95);
   \draw[very thick,->] (0,0.95) -- (0.5,1);
   \draw[very thick] (0.5,1) -- (1,1.05);
   \draw[very thick,->] (0,1.05) -- (0.25,1.075);
   \draw[very thick,rounded corners,->] (0.25,1.075) -- (0.5,1.1) -- (0.5,1.5);
   \draw[very thick] (0.5,1.5) -- (0.5,2);
   \node at (0.4,0.3) {$g$};
\end{tikzpicture}
\caption{}
\end{subfigure}
\begin{subfigure}{0.5\textwidth}
\centering
\begin{tikzpicture}[scale=1.5]
   \draw (0,0) -- (1,0) -- (1,2) -- (0,2) -- (0,0);
   \node at (-0.1,2.1) {$\tau$};
   \draw[very thick,->] (0.45,0) -- (0.45,0.5);
   \draw[very thick,rounded corners,->] (0.45,0.5) -- (0.45,0.89) -- (0.75,0.95);
   \draw[very thick] (0.75,0.95) -- (1,1);
   \draw[very thick,->] (0.55,0) -- (0.55,0.4);
   \draw[very thick,rounded corners,->] (0.55,0.4) -- (0.55,0.81) -- (0.75,0.85);
   \draw[very thick] (0.75,0.85) -- (1,0.9);
   \draw[very thick,->] (0,0.9) -- (0.5,1);
   \draw[very thick] (0.5,1) -- (1,1.1);
   \draw[very thick,->] (0,1) -- (0.25,1.05);
   \draw[very thick,rounded corners,->] (0.25,1.05) -- (0.55,1.11) -- (0.55,1.7);
   \draw[very thick] (0.55,1.7) -- (0.55,2);
   \draw[very thick,->] (0,1.1) -- (0.25,1.15);
   \draw[very thick,rounded corners,->] (0.25,1.15) -- (0.45,1.19) -- (0.45,1.6);
   \draw[very thick] (0.45,1.6) -- (0.45,2);
   \node at (0.35,0.3) {$g$};
\end{tikzpicture}
\caption{}
\end{subfigure}
\caption{In figure (a) we have a representation of $Z_{g,1}(\tau,\btau)$, while the dashed lines indicate an alternative fundamental domain corresponding to $\tau'=\tau-2$.  In (b) we focus on this new fundamental domain, paying attention to where the topological defect labeled by $g$ is located.  By deforming this picture we arrive at figure (c), which we recognize as a representation of $Z_{g,g^2}$.  For the group $\Z$ generated by an element $g$, any partial trace can be represented uniquely as such a diagram with $g$ lines that never cross.  For example, $Z_{2,3}$ is illustrated.}
\label{fig:TDLModOrbits}
\end{figure}

By taking the full set of modular transformations, we can start with untwisted sector partial traces $Z_{1,g^r}(\tau,\btau)$ and generate any partial trace $Z_{g^m,g^n}$ for any $m,n\in\Z$.  Some examples of how this works are shown in Figure \ref{fig:TDLModOrbits}.  If $G$ is cyclic, generated by $g$, and if we extend this to a $\Z$ action on our theory, then we get a unique picture for $Z_{m,n}(\tau,\btau)$ for each $m,n\in\Z$ in which we have only continuous $g$ lines which do not cross each other.

\subsection{Anomalies and partial trace periodicities}
\label{subsec:PTPeriodicities}

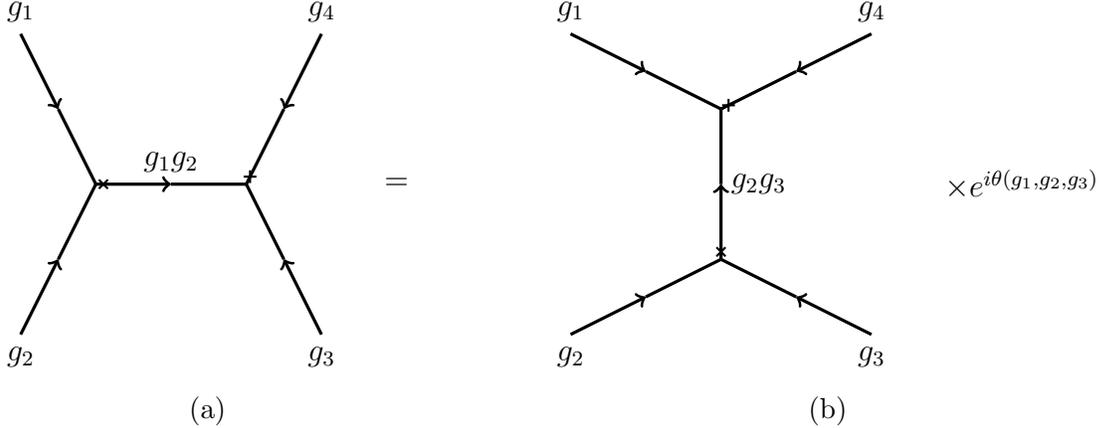
\begin{figure}
\begin{subfigure}{0.4\textwidth}
\centering
\begin{tikzpicture}[scale=1]
   \draw[very thick,->] (-2,-2) -- (-1.5,-1);
   \draw[very thick] (-1.5,-1) -- (-1,0);
   \draw[very thick,->] (-2,2) -- (-1.5,1);
   \draw[very thick] (-1.5,1) -- (-1,0);
   \draw[very thick,->] (-1,0) -- (0,0);
   \draw[very thick] (0,0) -- (1,0);
   \draw[very thick,->] (2,2) -- (1.5,1);
   \draw[very thick] (1.5,1) -- (1,0);
   \draw[very thick,->] (2,-2) -- (1.5,-1);
   \draw[very thick] (1.5,-1) -- (1,0);
   \draw[thick] (-0.96,-0.06) -- (-0.84,0.06);
   \draw[thick] (-0.96,0.06) -- (-0.84,-0.06);
   \draw[thick] (0.965,0.1) -- (1.135,0.1);
   \draw[thick] (1.05,0.015) -- (1.05,0.185);
   \node at (-2,2.3) {$g_1$};
   \node at (-2,-2.3) {$g_2$};
   \node at (2,-2.3) {$g_3$};
   \node at (2,2.3) {$g_4$};
   \node at (0,0.3) {$g_1g_2$};
   \node at (3,0) {$=$};
\end{tikzpicture}
\caption{}
\end{subfigure}
\begin{subfigure}{0.6\textwidth}
\centering
\begin{tikzpicture}[scale=1]
   \draw[very thick,->] (-2,-2) -- (-1,-1.5);
   \draw[very thick] (-1,-1.5) -- (0,-1);
   \draw[very thick,->] (-2,2) -- (-1,1.5);
   \draw[very thick] (-1,1.5) -- (0,1);
   \draw[very thick,->] (2,-2) -- (1,-1.5);
   \draw[very thick] (1,-1.5) -- (0,-1);
   \draw[very thick,->] (2,2) -- (1,1.5);
   \draw[very thick] (1,1.5) -- (0,1);
   \draw[very thick,->] (0,-1) -- (0,0);
   \draw[very thick] (0,0) -- (0,1);
   \draw[thick] (-0.06,-0.96) -- (0.06,-0.84);
   \draw[thick] (-0.06,-0.84) -- (0.06,-0.96);
   \draw[thick] (0.1,0.965) -- (0.1,1.135);
   \draw[thick] (0.015,1.05) -- (0.185,1.05);
   \node at (-2,2.3) {$g_1$};
   \node at (-2,-2.3) {$g_2$};
   \node at (2,-2.3) {$g_3$};
   \node at (2,2.3) {$g_4$};
   \node at (0.5,0) {$g_2g_3$};
   \node at (4,0) {$\times e^{i\theta(g_1,g_2,g_3)}$};
\end{tikzpicture}
\caption{}
\end{subfigure}
\caption{Three TDLs can be joined at a trivalent junction.  In full generality, an ordering should be specified, with the incoming lines listed in clockwork order; we mark the last line listed with a $\times$, following~\cite{Chang:2018iay}.  There are two ways of joining four TDLs (with $g_1g_2g_3g_4=1$) using trivalent junctions, and these can be related by a phase $\theta(g_1,g_2,g_3)$.  Inequivalent phases are classified by $H^3(G,\U(1))$.  If there is a representative for which the phase is trivial, then we say that the symmetry is non-anomalous.  Otherwise, the anomaly is given by the class in $H^3(G,\U(1))$.}
\label{fig:JunctionResolution}
\end{figure}

In this language, a network of these TDLs represents the coupling of the theory to a background gauge field (i.e.\ a flat connection) for the global symmetry group $G$.  If we wanted to gauge $G$, we would want to formally sum over all possible background configurations.  In order to do this consistently, preserving the topological nature of our defect lines, we need to be able to rearrange our network of lines~\cite{Gaiotto:2014kfa}.  In particular, if four TDLs come together in a junction (and the product of the four group elements must be the identity), then there are two ways to resolve this into a pair of trivalent junctions, and these two ways must be equivalent (up to a junction-dependent phase rotation).  The obstruction to this is measured by an element of $H^3(G,\U(1))$, the 't Hooft anomaly associated to $G$.  It can be captured by a phase relating the two different resolutions, as in Figure \ref{fig:JunctionResolution}.

For finite cyclic groups, the third cohomology is given by $H^3(\Z_N,\U(1))\cong\Z_N$.  The possible elements labeled by $0\le k<N$ have representatives given by~\cite{Bhardwaj:2017xup}
\be
\label{eq:AnomalyRep}
\theta(a,b,c)=2\pi ka\lp b+c-\langle b+c\rangle\rp/N^2,
\ee
where $\langle x\rangle$ is the mod $N$ representative of $x$ in the range $0\le x<N$ (so in particular it means that $\theta$ is a multiple of $2\pi/N$).

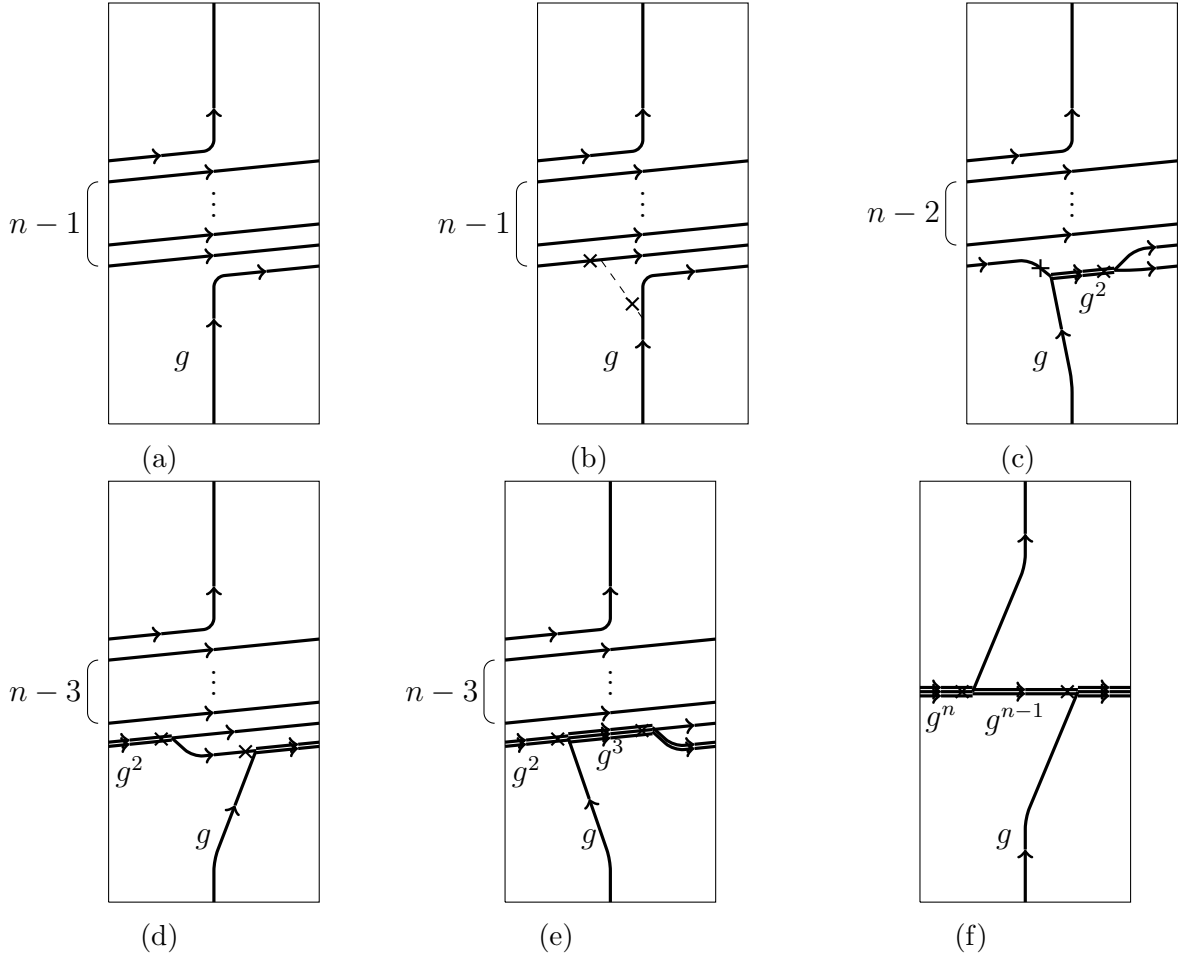
\begin{figure}
\begin{subfigure}{0.3\textwidth}
\centering
\begin{tikzpicture}[scale=2.8]
   \draw (0,0) -- (1,0) -- (1,2) -- (0,2) -- (0,0);
   \draw[very thick,->] (0.5,0) -- (0.5,0.5);
   \draw[very thick,rounded corners,->] (0.5,0.5) -- (0.5,0.7) -- (0.75,0.725);
   \draw[very thick] (0.75,0.725) -- (1,0.75);
   \draw[very thick,->] (0,0.75) -- (0.5,0.8);
   \draw[very thick] (0.5,0.8) -- (1,0.85);
   \draw[very thick,->] (0,0.85) -- (0.5,0.9);
   \draw[very thick] (0.5,0.9) -- (1,0.95);
   \draw[very thick,->] (0,1.15) -- (0.5,1.2);
   \draw[very thick] (0.5,1.2) -- (1,1.25);
   \draw[very thick,->] (0,1.25) -- (0.25,1.275);
   \draw[very thick,rounded corners,->] (0.25,1.275) -- (0.5,1.3) -- (0.5,1.5);
   \draw[very thick] (0.5,1.5) -- (0.5,2);
   \node at (0.5,1.08) {$\vdots$};
   \draw[rounded corners] (-0.05,0.75) -- (-0.1,0.75) -- (-0.1,1.15) -- (-0.05,1.15);
   \node at (-0.3,0.95) {$n-1$};
%   \draw[decorate,decoration={brace,amplitude=10pt,raise=4pt}] (0,0.75) -- (0,1.15) node [midway,xshift=-1cm] {$n-1$};
   \node at (0.35,0.3) {$g$};
\end{tikzpicture}
\caption{}
\end{subfigure}
\begin{subfigure}{0.3\textwidth}
\centering
\begin{tikzpicture}[scale=2.8]
   \draw (0,0) -- (1,0) -- (1,2) -- (0,2) -- (0,0);
   \draw[very thick,->] (0.5,0) -- (0.5,0.4);
   \draw[very thick,rounded corners,->] (0.5,0.4) -- (0.5,0.7) -- (0.75,0.725);
   \draw[very thick] (0.75,0.725) -- (1,0.75);
   \draw[very thick,->] (0,0.75) -- (0.5,0.8);
   \draw[very thick] (0.5,0.8) -- (1,0.85);
   \draw[very thick,->] (0,0.85) -- (0.5,0.9);
   \draw[very thick] (0.5,0.9) -- (1,0.95);
   \draw[very thick,->] (0,1.15) -- (0.5,1.2);
   \draw[very thick] (0.5,1.2) -- (1,1.25);
   \draw[very thick,->] (0,1.25) -- (0.25,1.275);
   \draw[very thick,rounded corners,->] (0.25,1.275) -- (0.5,1.3) -- (0.5,1.5);
   \draw[very thick] (0.5,1.5) -- (0.5,2);
   \node at (0.5,1.08) {$\vdots$};
   \draw[dashed] (0.5,0.5) -- (0.3,0.78);
   \draw[thick] (0.42,0.6) -- (0.48,0.54);
   \draw[thick] (0.42,0.54) -- (0.48,0.6);
   \draw[thick] (0.22,0.745) -- (0.28,0.805);
   \draw[thick] (0.22,0.805) -- (0.28,0.745);
   \draw[rounded corners] (-0.05,0.75) -- (-0.1,0.75) -- (-0.1,1.15) -- (-0.05,1.15);
   \node at (-0.3,0.95) {$n-1$};
%   \draw[decorate,decoration={brace,amplitude=10pt,raise=4pt}] (0,0.75) -- (0,1.15) node [midway,xshift=-1cm] {$n-1$};
   \node at (0.35,0.3) {$g$};
\end{tikzpicture}
\caption{}
\end{subfigure}
\begin{subfigure}{0.3\textwidth}
\centering
\begin{tikzpicture}[scale=2.8]
   \draw (0,0) -- (1,0) -- (1,2) -- (0,2) -- (0,0);
   \draw[very thick,rounded corners,->] (0.5,0) -- (0.5,0.2) -- (0.45,0.45);
   \draw[very thick] (0.45,0.45) -- (0.4,0.7);
%   \draw[very thick,->] (0.5,0) -- (0.5,0.2);
%   \draw[very thick,rounded corners,->] (0.5,0.2) -- (0.5,0.7) -- (0.75,0.725);
%   \draw[very thick] (0.75,0.725) -- (1,0.75);
%   \draw[very thick,->] (0,0.75) -- (0.5,0.8);
%   \draw[very thick] (0.5,0.8) -- (1,0.85);
   \draw[very thick,->] (0,0.85) -- (0.5,0.9);
   \draw[very thick] (0.5,0.9) -- (1,0.95);
   \draw[very thick,->] (0,1.15) -- (0.5,1.2);
   \draw[very thick] (0.5,1.2) -- (1,1.25);
   \draw[very thick,->] (0,1.25) -- (0.25,1.275);
   \draw[very thick,rounded corners,->] (0.25,1.275) -- (0.5,1.3) -- (0.5,1.5);
   \draw[very thick] (0.5,1.5) -- (0.5,2);
   \node at (0.5,1.08) {$\vdots$};
   \draw[very thick,->] (0,0.75) -- (0.1,0.76);
   \draw[very thick,rounded corners] (0.1,0.76) -- (0.3,0.78) -- (0.4,0.7);
   \draw[thick] (0.35,0.6976) -- (0.35,0.7824);
   \draw[thick] (0.3076,0.74) -- (0.3924,0.74);
   \draw[very thick,->] (0.4,0.71) -- (0.55,0.725);
   \draw[very thick] (0.55,0.725) -- (0.7,0.74);
   \draw[very thick,->] (0.4,0.69) -- (0.55,0.705);
   \draw[very thick] (0.55,0.705) -- (0.7,0.72);
   \draw[thick] (0.62,0.755) -- (0.68,0.695);
   \draw[thick] (0.62,0.695) -- (0.68,0.75);
   \draw[very thick,rounded corners,->] (0.7,0.73) -- (0.8,0.83) -- (0.9,0.84);
   \draw[very thick] (0.9,0.84) -- (1,0.85);
   \draw[very thick,rounded corners,->] (0.7,0.73) -- (0.8,0.73) -- (0.9,0.74);
   \draw[very thick] (0.9,0.74) -- (1,0.75);
   \draw[rounded corners] (-0.05,0.85) -- (-0.1,0.85) -- (-0.1,1.15) -- (-0.05,1.15);
   \node at (-0.3,1) {$n-2$};
%   \draw[decorate,decoration={brace,amplitude=10pt,raise=4pt}] (0,0.85) -- (0,1.15) node [midway,xshift=-1cm] {$n-2$};
   \node at (0.35,0.3) {$g$};
   \node at (0.6,0.6) {$g^2$};
\end{tikzpicture}
\caption{}
\end{subfigure}
\begin{subfigure}{0.3\textwidth}
\centering
\begin{tikzpicture}[scale=2.8]
   \draw (0,0) -- (1,0) -- (1,2) -- (0,2) -- (0,0);
   \draw[very thick,rounded corners,->] (0.5,0) -- (0.5,0.2) -- (0.6,0.46);
   \draw[very thick] (0.6,0.46) -- (0.7,0.72);
   \draw[very thick,->] (0,0.85) -- (0.5,0.9);
   \draw[very thick] (0.5,0.9) -- (1,0.95);
   \draw[very thick,->] (0,1.15) -- (0.5,1.2);
   \draw[very thick] (0.5,1.2) -- (1,1.25);
   \draw[very thick,->] (0,1.25) -- (0.25,1.275);
   \draw[very thick,rounded corners,->] (0.25,1.275) -- (0.5,1.3) -- (0.5,1.5);
   \draw[very thick] (0.5,1.5) -- (0.5,2);
   \node at (0.5,1.08) {$\vdots$};
   \draw[very thick,->] (0,0.76) -- (0.1,0.77);
   \draw[very thick,->] (0,0.74) -- (0.1,0.75);
   \draw[very thick] (0.1,0.77) -- (0.3,0.79);
   \draw[very thick] (0.1,0.75) -- (0.3,0.77);
   \draw[thick] (0.22,0.745) -- (0.28,0.805);
   \draw[thick] (0.22,0.805) -- (0.28,0.745);
   \draw[very thick,->] (0.3,0.78) -- (0.6,0.81);
   \draw[very thick] (0.6,0.81) -- (1,0.85);
   \draw[very thick,rounded corners,->] (0.3,0.78) -- (0.4,0.69) -- (0.5,0.7);
   \draw[very thick] (0.5,0.7) -- (0.7,0.72);
   \draw[thick] (0.62,0.685) -- (0.68,0.745);
   \draw[thick] (0.62,0.745) -- (0.68,0.685);
   \draw[very thick,->] (0.7,0.73) -- (0.9,0.75);
   \draw[very thick] (0.9,0.75) -- (1,0.76);
   \draw[very thick,->] (0.7,0.71) -- (0.9,0.73);
   \draw[very thick] (0.9,0.73) -- (1,0.74);
   \draw[rounded corners] (-0.05,0.85) -- (-0.1,0.85) -- (-0.1,1.15) -- (-0.05,1.15);
   \node at (-0.3,1) {$n-3$};
   \node at (0.45,0.3) {$g$};
   \node at (0.1,0.62) {$g^2$};
\end{tikzpicture}
\caption{}
\end{subfigure}
\begin{subfigure}{0.33\textwidth}
\centering
\begin{tikzpicture}[scale=2.8]
   \draw (0,0) -- (1,0) -- (1,2) -- (0,2) -- (0,0);
   \draw[very thick,rounded corners,->] (0.5,0) -- (0.5,0.2) -- (0.4,0.49);
   \draw[very thick] (0.4,0.49) -- (0.3,0.78);
   \draw[very thick,->] (0,0.85) -- (0.5,0.9);
   \draw[very thick] (0.5,0.9) -- (1,0.95);
   \draw[very thick,->] (0,1.15) -- (0.5,1.2);
   \draw[very thick] (0.5,1.2) -- (1,1.25);
   \draw[very thick,->] (0,1.25) -- (0.25,1.275);
   \draw[very thick,rounded corners,->] (0.25,1.275) -- (0.5,1.3) -- (0.5,1.5);
   \draw[very thick] (0.5,1.5) -- (0.5,2);
   \node at (0.5,1.08) {$\vdots$};
   \draw[very thick,->] (0,0.76) -- (0.1,0.77);
   \draw[very thick,->] (0,0.74) -- (0.1,0.75);
   \draw[very thick] (0.1,0.77) -- (0.3,0.79);
   \draw[very thick] (0.1,0.75) -- (0.3,0.77);
   \draw[thick] (0.22,0.745) -- (0.28,0.805);
   \draw[thick] (0.22,0.805) -- (0.28,0.745);
   \draw[very thick,->] (0.3,0.76) -- (0.5,0.78);
   \draw[very thick,->] (0.3,0.78) -- (0.5,0.8);
   \draw[very thick,->] (0.3,0.8) -- (0.5,0.82);
   \draw[very thick] (0.5,0.78) -- (0.7,0.8);
   \draw[very thick] (0.5,0.8) -- (0.7,0.82);
   \draw[very thick] (0.5,0.82) -- (0.7,0.84);
   \draw[thick] (0.62,0.785) -- (0.68,0.845);
   \draw[thick] (0.62,0.845) -- (0.68,0.785);
   \draw[very thick,->] (0.7,0.82) -- (0.9,0.84);
   \draw[very thick] (0.9,0.84) -- (1,0.85);
   \draw[very thick,rounded corners,->] (0.7,0.81) -- (0.8,0.72) -- (0.9,0.73);
   \draw[very thick,rounded corners,->] (0.7,0.83) -- (0.8,0.74) -- (0.9,0.75);
%   \draw[very thick,rounded corners,->] (0.3,0.78) -- (0.4,0.69) -- (0.5,0.7);
%   \draw[very thick] (0.5,0.7) -- (0.7,0.72);
%   \draw[thick] (0.62,0.685) -- (0.68,0.745);
%   \draw[thick] (0.62,0.745) -- (0.68,0.685);
%   \draw[very thick,->] (0.7,0.73) -- (0.9,0.75);
   \draw[very thick] (0.9,0.75) -- (1,0.76);
%   \draw[very thick,->] (0.7,0.71) -- (0.9,0.73);
   \draw[very thick] (0.9,0.73) -- (1,0.74);
   \draw[rounded corners] (-0.05,0.85) -- (-0.1,0.85) -- (-0.1,1.15) -- (-0.05,1.15);
   \node at (-0.3,1) {$n-3$};
   \node at (0.4,0.3) {$g$};
   \node at (0.1,0.62) {$g^2$};
   \node at (0.5,0.7) {$g^3$};
\end{tikzpicture}
\caption{}
\end{subfigure}
\begin{subfigure}{0.33\textwidth}
\centering
\begin{tikzpicture}[scale=2.8]
   \draw (0,0) -- (1,0) -- (1,2) -- (0,2) -- (0,0);
   \draw[very thick,->] (0,0.98) -- (0.1,0.98);
   \draw[very thick,->] (0,1) -- (0.1,1);
   \draw[very thick,->] (0,1.02) -- (0.1,1.02);
   \draw[very thick] (0.1,0.98) -- (0.25,0.98);
   \draw[very thick] (0.1,1) -- (0.25,1);
   \draw[very thick] (0.1,1.02) -- (0.25,1.02);
   \draw[very thick,->] (0.25,0.99) -- (0.5,0.99);
   \draw[very thick,->] (0.25,1.01) -- (0.5,1.01);
   \draw[very thick] (0.5,0.99) -- (0.75,0.99);
   \draw[very thick] (0.5,1.01) -- (0.75,1.01);
   \draw[very thick,->] (0.75,0.98) -- (0.9,0.98);
   \draw[very thick,->] (0.75,1) -- (0.9,1);
   \draw[very thick,->] (0.75,1.02) -- (0.9,1.02);
   \draw[very thick] (0.9,0.98) -- (1,0.98);
   \draw[very thick] (0.9,1) -- (1,1);
   \draw[very thick] (0.9,1.02) -- (1,1.02);
   \draw[very thick,->] (0.5,0) -- (0.5,0.25);
   \draw[very thick,rounded corners] (0.5,0.25) -- (0.5,0.4) -- (0.75,1);
   \draw[very thick,rounded corners,->] (0.25,1) -- (0.5,1.6) -- (0.5,1.75);
   \draw[very thick] (0.5,1.75) -- (0.5,2);
   \draw[thick] (0.17,0.97) -- (0.23,1.03);
   \draw[thick] (0.17,1.03) -- (0.23,0.97);
   \draw[thick] (0.67,0.97) -- (0.73,1.03);
   \draw[thick] (0.67,1.03) -- (0.73,0.97);
   \node at (0.4,0.3) {$g$};
   \node at (0.1,0.87) {$g^n$};
   \node at (0.45,0.87) {$g^{n-1}$};
   \node at (-0.3,1) {$\hphantom{n-3}$};
\end{tikzpicture}
\caption{}
\end{subfigure}
\caption{We start with the $Z_{g,g^n}$ partial trace in (a).  In (b) we draw a dashed identity line between two $g$ lines.  Going to (c), we use a crossing relation, which results in a $g^2$ line connecting the two junctions, and which multiplies the evaluation of the diagram by a phase $e^{i\theta(g^{-1},g,g)}$, where $\theta\in H^3(G,\U(1))$ represents the anomaly.  From (c) to (d) we simply deform the picture, sliding the junction on the right around the cycle on the torus.  From (d) to (e) we perform another crossing, introducing an additional factor $e^{i\theta(g^{-1},g^2,g)}$.  Repeating the sliding and crossing steps until all the horizontal $g$ lines have been absorbed we get to (f).  The total phase from the crossings is $\g=\exp(i\sum_{j=1}^{n-1}\theta(g^{-1},g^j,g))$.}
\label{fig:PTPeriodicity}
\end{figure}

In the case that $G\cong\Z_N$, with generator $g$ (with anomaly given by the element $k\in H^3(\Z_N,\U(1))\cong\Z_N$), then we can always replace a $g^N$ line with an identity line which can be erased.  On the other hand, we can always uniquely generate a candidate $Z_{m,n}$ partial trace by applying modular transformations to the untwisted sector.  What then is the relation between partial traces whose subscripts differ by multiples of $N$?  It turns out that these will be related by a phase $\g$ which will be an $N$th root of unity.  A general argument for this is given in section 4.4 of~\cite{Chang:2018iay}, but it can also be obtained by explicit manipulations.  As an example, in Figure~\ref{fig:PTPeriodicity}, we exhibit steps that can be used relate $Z_{1,0}$ with $Z_{1,N}$.  The net phase that accumulates through these steps is
\be
\g=\exp\lp i\sum_{j=1}^{n-1}\theta(g^{-1},g^j,g)\rp=\exp\lp -2\pi ik/N\rp,
\ee
where we have used the fact that (for the representative (\ref{eq:AnomalyRep}) above) $\theta(-1,j,1)=0$ for $j<N-1$, and $\theta(-1,N-1,1)=-2\pi k/N$.  Similar manipulations can show that shifting a partial trace subscript by a multiple of $N$ will always multiply it by an $N$th root of unity, and thus also that the partial traces are $N^2$-periodic in their subscripts.

%\subsection{Generalities}

\section{Continuous symmetries}
\label{sec:Flavored}

\subsection{Flavored partition function}

In the case that our theory has a continuous symmetry, we can introduce extra machinery which will ultimately simplify orbifold calculations.  Assuming we can express the partition function as a trace over states in the theory, define the \textit{flavored} partition function as

\be
\label{zf}
Z^f(\tau,\bar{\tau};z_L,z_R)=\text{Tr}\left[q^{L_0-\frac{c}{24}}\bar{q}^{\bar{L}_0-\frac{\bar{c}}{24}}e^{2\pi iz_LJ_L}e^{-2\pi iz_RJ_R}\right],
\ee
where $J_L$ and $J_R$ are the zero modes of left- and right-moving currents that generate continuous symmetries of our theory. From path integral arguments, $Z^f$ is expected to have a theory-independent modular transformation law \cite{Kraus:2006wn},

\be
\label{zftransform}
Z^f\left(\frac{a\tau+b}{c\tau+d},\frac{a\bar{\tau}+b}{c\bar{\tau}+d};z_L,z_R\right)=e^{\pi ik\left(c(c\tau+d)z_L^2-c(c\bar{\tau}+d)z_R^2\right)}Z^f(\tau,\bar{\tau};(c\tau+d)z_L,(c\bar{\tau}+d)z_R).
\ee

Here $k$ is a constant that will depend on our chosen conventions for normalization of the currents.  It shows up in the current-current OPE as
\be
J(w)J(0)\sim\frac{k}{w^2}+\cdots,
\ee
or equivalently in the mode algebra, $[J_m,J_n]=km\d_{m+n,0}$, where $J_m=\oint\tfrac{dw}{2\pi i}z^mJ(w)$ are the modes of the current $J(w)$ (and similarly for the right-moving current), and where $J_L$ is simply $J_0$.  For abelian currents (which is all we are presently concerned with), the number $k$ can always be rescaled by changing the definitions of the currents.  The formulas above won't change provided we also rescale $z_L$ and $z_R$; only the combination $kz^2$ is physical.
%For the free boson it equivalently shows up in the commutation relation $[\alpha_n,\alpha_m]=n\frac{k}{2}\delta_{nm}$, so for those currents we will have $k=2$.

For the free boson (using the conventions of \cite{Polchinski:1998rq} with $\al'=1$), if we want to identify $J_L$ with $p_L$ and $J_R$ with $p_R$, then we would need to take $J(z)\sim 2i\p X(z)$, and from the $JJ$ OPE we would read off $k=2$.  Similarly identifying $J_R$ with $p_R$, the flavored partition function takes the form
%In that case, we have a U(1) symmetry on the holomorphic and antiholomorphic sides.  The zero modes of the corresponding current are the left and right momenta, so the corresponding flavored partition function takes the form
\be
Z_R(\tau,\btau;z_L,z_R)=|\eta(\tau)|^{-2}\sum_{n,w\in\Z}y_L^{p_L}y_R^{p_R}q^{\frac{p_L^2}{4}}\bq^{\frac{p_R^2}{4}}
\ee
with $y=e^{2\pi iz}$.  One can explicitly verify (\ref{zftransform}) for the boson by Poisson resummation, which in turn verifies it for any set of U(1) currents \cite{Zamolodchikov85}, which will be the main objects of interest in this paper.  It is the existence of a reliable, known behavior under modular transformations that will allow flavored partition functions to aid our calculation of orbifold partition functions.

\subsection{Implementing the method of modular orbits}
\label{internaltheory}

If, for some specific $\alpha_L$, $\alpha_R$, the element $g=\text{exp}[2\pi i(\alpha_L J_L-\alpha_R J_R)]$ generates a cyclic symmetry group of our states, we can easily compute the partition function resulting from orbifolding by $g$.  The untwisted partition function with $r$ insertions of $g$ is given by $Z_{1,r}(\tau,\btau)=Z^f(\tau,\btau;r\alpha_L,r\alpha_R)$.  The modular transformation (\ref{zftransform}) of $Z^f$ then allows us to calculate all partial traces in terms of the flavored partition function; this combined with (\ref{eq:ModTransOrbTraces}) gives $Z_{m,n}$ as
\bea
\label{eq:zmnflavor}
Z_{m,n}(\tau,\btau) &=& e^{\pi ik\left(m^2(\tau\alpha_L^2-\bar{\tau}\alpha_R^2)-mn(\alpha_L^2-\alpha_R^2)\right)}Z^f(\tau,\btau;(n-m\tau)\alpha_L,(n-m\bar{\tau})\alpha_R)\\
&=& \sum_{\mathrm{states\ }i}e^{2\pi in\ls\al_LQ_i-\al_R\bar{Q}_i-\frac{km}{2}\lp\al_L^2-\al_R^2\rp\rs} q^{h_i-m\al_LQ_i+\frac{km^2}{2}\al_L^2-\frac{c}{24}}\bq^{\bar{h}_i-m\al_R\bar{Q}_i+\frac{km^2}{2}\al_R^2-\frac{\bar{c}}{24}},\non
\eea
where $Q_i$ and $\bar{Q}_i$ are the eigenvalues of $J_L$ and $J_R$ on the state $i$ of the parent theory.

Using (\ref{eq:zmnflavor}), it's now straightforward to write the partition function as a sum over the states of the parent theory.  The result is that the partition function in the $m$-twisted sector should be
\be
\label{zm}
Z_m=\frac{1}{|\Z|}\sum_{\substack{n\in\Z \\ \text{states } i}}e^{2\pi in\left[\alpha_LQ_i-\alpha_R\bar{Q}_i-\frac{km}{2}(\alpha_L^2-\alpha_R^2)\right]}q^{h_i-m\alpha_LQ_i+\frac{km^2}{2}\alpha_L^2-\frac{c}{24}}\bar{q}^{\bar{h}_i-m\alpha_R\bar{Q}_i+\frac{km^2}{2}\alpha_R^2-\frac{\bar{c}}{24}}.
\ee

If $g$ is in fact order $N$ in the parent theory, then as shown in section~\ref{subsec:PTPeriodicities}, the partial traces will actually be at worst $N^2$ periodic.  In this case, then we can dispense with the infinite $|\Z|$ constant in (\ref{zm}) and write
\be
\label{eq:zmRenorm}
Z_m=\frac{1}{N^2}\sum_{n=0}^{N^2-1}\sum_{\text{states } i}e^{2\pi in\left[\alpha_LQ_i-\alpha_R\bar{Q}_i-\frac{km}{2}(\alpha_L^2-\alpha_R^2)\right]}q^{h_i-m\alpha_LQ_i+\frac{km^2}{2}\alpha_L^2-\frac{c}{24}}\bar{q}^{\bar{h}_i-m\alpha_R\bar{Q}_i+\frac{km^2}{2}\alpha_R^2-\frac{\bar{c}}{24}}.
\ee

\subsection{Interpretation as a modified group projection in twisted sectors}

Note that the term
\be
\frac{1}{|\Z|}\sum_{n\in\Z}e^{2\pi in\left[\alpha_LJ_L-\alpha_RJ_R-\frac{km}{2}(\alpha_L^2-\alpha_R^2)\right]}\qquad\mathrm{or}\qquad\frac{1}{N^2}\sum_{n=0}^{N^2-1}e^{2\pi in\left[\alpha_LJ_L-\alpha_RJ_R-\frac{km}{2}(\alpha_L^2-\alpha_R^2)\right]}
\ee
in (\ref{zm}) has the form of a projection operator, forcing
\be
\label{proj}
\alpha_LJ_L-\alpha_RJ_R-\frac{km}{2}(\alpha_L^2-\alpha_R^2)\in\Z
\ee
on all states.  It is this projection that will keep our resultant theory modular invariant; it is the analog of level matching in the usual orbifold prescription.  To better understand its effect on states, we can rewrite the above condition as (where $\rho(g)$ represents the action of $g$ on the Hilbert space)
\be
\label{projstate}
e^{2\pi i(\alpha_LJ_L-\alpha_RJ_R)}\ket{\psi}=\rho(g)\ket{\psi}=e^{2\pi im\frac{k}{2}(\alpha_L^2-\alpha_R^2)}\ket{\psi}.
\ee
When $\alpha_L^2=\alpha_R^2$, this has the familiar role of projecting onto $\alpha_LJ_L-\alpha_RJ_R\in\Z$ i.e. projecting onto group-invariant states.  When $\alpha_L^2\neq\alpha_R^2$, as is generically the case for asymmetric actions, the projection gets modified.  There is now a sector-dependent phase which tells us how the twisted sectors match up with each other.  Furthermore, as we will see below, it will tell us the lowest power of $g$ for which the twisted sectors are generically non-empty, indicating that we have effectively quotiented by this power of $g$, which may now act symmetrically on the parent theory.  In this way we may find that our asymmetric orbifold was equivalent to a symmetric one.

With this in mind, we can write the partition traces in the individual twisted sectors as
\be
\label{generalzm}
Z_m=\sum_{\substack{\text{states} \\ \text{satisfying} \\ (\ref{proj})}}q^{h_i-m\alpha_LJ_L+\frac{km^2}{2}\alpha_L^2-\frac{c}{24}}\bar{q}^{\bar{h}_i-m\alpha_RJ_R+\frac{km^2}{2}\alpha_R^2-\frac{\bar{c}}{24}}.
\ee

\section{Free boson examples}
\label{sec:Boson}

\subsection{Conventions}

Let's apply these results to the specific case of the compact free boson, with the conventions introduced in section~\ref{sec:examples}.  That theory has (anti-)holomorphic U(1) currents with zero mode $p_{L,R}$, so a natural choice would then be $J_L=p_L$, $J_R=p_R$.  As previously determined, our conventions dictate that $k=2$ for this choice of currents.  Since the weights of states in that theory are also given in terms of momenta, our expression (\ref{generalzm}) for $Z_m$ simplifies to
\be
Z_m=|\eta|^{-2}\sum_{\substack{\text{states} \\ \text{satisfying} \\ (\ref{proj})}}q^{\frac{1}{4}(p_L-2m\alpha_L)^2}\bar{q}^{\frac{1}{4}(p_R-2m\alpha_R)^2}.
\ee
We'll express the momenta in the usual way, $p_{L,R}=x/R\pm yR$.  We would like to regard our group elements $g=\text{exp}[2\pi i(\alpha_L J_L-\alpha_R J_R)]$ as being built from the integers $x$ and $y$.  Then a natural choice for $\alpha$ will be
\be
\alpha_L=\frac{1}{2N}\left(\frac{\beta}{R}+\gamma R\right),\hspace{.5cm} \alpha_R=\frac{1}{2N}\left(\frac{\beta}{R}-\gamma R\right)
\ee
where $N\in\Z^{>0},(\beta,\gamma)\in(\Z)^2$ and we stipulate that gcd$(\beta,\gamma,N)=1$.  This choice gives us the group action
\be
g=e^{2\pi i(\beta y+\gamma x)/N}
\ee
which means we can build both symmetric and asymmetric $\Z_N$ actions out of arbitrary combinations of $x$ and $y$ at any radius.  Given these choices, the partition traces (\ref{generalzm}) are
\be
\label{bosonz}
Z_m=|\eta|^{-2}\sum q^{\frac{1}{4}\left[\frac{1}{R}\left(x-\frac{m\beta}{N}\right)+R\left(y-\frac{m\gamma}{N}\right)\right]^2}\bar{q}^{\frac{1}{4}\left[\frac{1}{R}\left(x-\frac{m\beta}{N}\right)-R\left(y-\frac{m\gamma}{N}\right)\right]^2}
\ee
%and the projection (\ref{proj}) becomes
where the sum is over integers $x$ and $y$ satisfying
\be
\label{gammabetaproj}
%e^{2\pi i(\beta y+\gamma x)/N}\ket{x,y}=e^{2\pi im\beta\gamma/N^2}\ket{x,y}.
\frac{\beta y+\gamma x}{N}-\frac{ m\beta\gamma}{N^2}\in\Z.
\ee
In this language, symmetric actions correspond to the case $\beta=0$, and the case $\gamma=0$ corresponds to the T-dual of a symmetric action.  These are shifts by the coordinate or dual coordinate, respectively, on the circle.  Note again that it is precisely this case in which the condition (\ref{gammabetaproj}) becomes group invariance (rather than covariance).

In order to obtain the full partition function we need to know which values of $m$ to sum over.  The default assumption for a symmetric orbifold is that we would take $m\in\Z_N$, perform the sum and get a good result.  However, that may not be compatible with modular invariance.  Luckily, the projection can guide us.  (\ref{gammabetaproj}) implies that for a sector to be nonempty, we need values of $x$ and $y$ satisfying
\be
\gamma x+\beta y-\frac{m\beta\gamma}{N}\in N\Z.
\ee
A necessary condition is then that $m$ be a multiple of $N/\text{gcd}(\beta\gamma,N)$.  At the point that $m$ becomes a multiple of $N^2/\text{gcd}(\beta\gamma,N)$, the term $m\beta\gamma/N$ is a multiple of $N$ and can be absorbed into the right-hand side, i.e.\ the corresponding solutions for $x$ and $y$ are exactly the same as in the untwisted sector $m=0$.  Similar arguments show that there are at most $N$ values of $m$ that produce distinct sets of solutions $x$ and $y$, and we should take
\be
\label{onebosonm}
m\in\frac{N}{\text{gcd}(\beta\gamma,N)}\Z_N.
\ee
Incorporating this into the partition function allows us to be explicit about our sum.  With $\rho\equiv\text{gcd}(\beta\gamma,N)$, the full orbifold partition function will be
\be
\label{eq:fullbosonorbifold}
|\eta|^{-2}\sum_{\substack{m'\in\Z_N \\ x,y\in\Z^2 \\ \gamma x+\beta y-\frac{\gamma\beta m'}{\rho}\in N\Z}}q^{\frac{1}{4}\left[\frac{1}{R}\left(x-\frac{m'\beta}{\rho}\right)+R\left(y-\frac{m'\gamma}{\rho}\right)\right]^2}\bar{q}^{\frac{1}{4}\left[\frac{1}{R}\left(x-\frac{m'\beta}{\rho}\right)-R\left(y-\frac{m'\gamma}{\rho}\right)\right]^2},
\ee
where we now sum over $m'=\rho m/N$.

\subsection{Examples}

\subsubsection{Single Bosons}

Let's check that this reproduces expected results.  We will recover the results of section \ref{sec:examples} by identifying the relevant values of $\beta, \gamma$ and $N$ and appealing to (\ref{eq:fullbosonorbifold}).  First take the very simple case $N=2, \beta=0, \gamma=1$; this is the $\Z_2$ coordinate shift which led to (\ref{z2coordshift}).  (\ref{eq:fullbosonorbifold}) is then
\be
|\eta|^{-2}\sum_{\substack{m\in\Z_2 \\ y\in\Z \\ x\in2\Z}}q^{\frac{1}{4}\left[\frac{x}{R}+R\left(y-\frac{m}{2}\right)\right]^2}\bar{q}^{\frac{1}{4}\left[\frac{x}{R}-R\left(y-\frac{m}{2}\right)\right]^2}=|\eta|^{-2}\sum_{x',y'\in\Z^2}q^{\frac{1}{4}\left[\frac{2x'}{R}+\frac{y'R}{2}\right]^2}\bar{q}^{\frac{1}{4}\left[\frac{2x'}{R}-\frac{y'R}{2}\right]^2}
\ee
which is, as we would expect, the same theory at a radius $R/2$.

The case $N=2, \beta=\gamma=1$ corresponds to the group element which was written as $g=(\pi,\pi;1)$ in section \ref{sec:examples}.  This is an asymmetric shift, given by a combination of the coordinate and its dual.  Our flavored formalism immediately yields the result
\begin{multline}
|\eta|^{-2}\sum_{\substack{m\in\Z_2 \\ x,y\in\Z^2 \\ x+y-m\in2\Z}}q^{\frac{1}{4}\left[\frac{1}{R}(x-m)+R(y-m)\right]^2}\bar{q}^{\frac{1}{4}\left[\frac{1}{R}(x-m)-R(y-m)\right]^2} \\
=|\eta|^{-2}\sum_{\substack{x,y\in\Z^2 \\ m\in\Z_2}}q^{\frac{1}{4}\left[\frac{1}{R}(x+y+m)+R(x-y)\right]^2}\bar{q}^{\frac{1}{4}\left[\frac{1}{R}(x+y+m)-R(x-y)\right]^2}.
\end{multline}
We note that the untwisted ($m=0$) sector contains states with $R$ and $1/R$ coefficients of like parity, while the twisted ($m=1$) sector contains those with opposite parity.  In total we have the same set of states, so the orbifold acted trivially\footnote{Of course in general we can't conclude that the parent theory and the orbifold theory are the same just because they have the same spectrum; we would need to know and compare their OPEs as well.  But for the case of $c=1$, since we know the full classification of consistent theories we know that knowledge of the spectrum is sufficient to reach this conclusion.
%In general we would need to examine the action on the OPEs as well to conclude that the theory hasn't changed, but the classification of $c=1$ theories allows us this shortcut.
}.  If we instead took $N=4, \beta=2, \gamma=1$, we would get the theory back at radius $R/2$.  Note, then, that both of these attempts at asymmetric orbifolds produced the same result as orbifolding by $g^2$ rather than $g$.  Does this result generalize?

We can check by slightly rearranging (\ref{eq:fullbosonorbifold}).  Write $m=m'+k\rho$ where $\rho=\gcd(N,\beta\g)$, $m'$ is valued in $\Z_\rho$ and $k$ is valued in $\Z_{N/\rho}$.  The partition function is then
\be
|\eta|^{-2}\sum_{\substack{m'\in\Z_\rho \\ k\in\Z_{N/\rho} \\ x,y\in\Z^2 \\ \gamma x+\beta y-\frac{\gamma\beta}{\rho}(m'+\rho k)\in N\Z}}q^{\frac{1}{4}\left[\frac{1}{R}\left(x-\frac{\beta}{\rho}(m'+\rho k)\right)+R\left(y-\frac{\gamma}{\rho}(m'+\rho k)\right)\right]^2}\bar{q}^{\frac{1}{4}\left[\frac{1}{R}\left(x-\frac{\beta}{\rho}(m'+\rho k)\right)-R\left(y-\frac{\gamma}{\rho}(m'+\rho k)\right)\right]^2}.
\ee
Shifting $x\to x+\beta k$ and $y\to y+\gamma k$ gives
\be
|\eta|^{-2}\sum_{\substack{m'\in\Z_\rho \\ k\in\Z_{N/\rho} \\ x,y\in\Z^2 \\ \gamma x+\beta y-\frac{m'\gamma\beta}{\rho}+k\gamma\beta\in N\Z}}q^{\frac{1}{4}\left[\frac{1}{R}\left(x-\frac{m'\beta}{\rho}\right)+R\left(y-\frac{m'\gamma}{\rho}\right)\right]^2}\bar{q}^{\frac{1}{4}\left[\frac{1}{R}\left(x-\frac{m'\beta}{\rho}\right)-R\left(y-\frac{m'\gamma}{\rho}\right)\right]^2}.
\ee
Here $k$ now appears only in the constraint, and can be reinterpreted as follows.  Since for any integers $x$ and $y$ satisfying
\be
\gamma x+\beta y-\frac{m'\gamma\beta}{\rho}\in\rho\Z
\ee
we can uniquely choose $k\in\Z_{N/\rho}$ so that
\be
\gamma x+\beta y-\frac{m'\gamma\beta}{\rho}+k\gamma\beta\in N\Z,
\ee
the partition function above can be simplified as
\be
|\eta|^{-2}\sum_{\substack{m'\in\Z_\rho \\ x,y\in\Z^2 \\ \gamma x+\beta y-\frac{m'\gamma\beta}{\rho}\in \rho\Z}}q^{\frac{1}{4}\left[\frac{1}{R}\left(x-\frac{m'\beta}{\rho}\right)+R\left(y-\frac{m'\gamma}{\rho}\right)\right]^2}\bar{q}^{\frac{1}{4}\left[\frac{1}{R}\left(x-\frac{m'\beta}{\rho}\right)-R\left(y-\frac{m'\gamma}{\rho}\right)\right]^2},
\ee
which is manifestly equivalent to the orbifold by the group generated by $g^{N/\rho}$.  As promised, this tells us the lowest power of $g$ by which we can effectively orbifold.
%With k appearing only in the constraint, its rhs becomes $N\Z-\beta\gamma\Z_{N/\rho}\cong\rho\Z$ (this follows from B\'{e}zout's identity) and this becomes equivalent to the orbifold by $g^{N/\rho}$.  As promised, then, the formalism lets us know the lowest power of $g$ by which we can consistently orbifold.  (Note that a trivial orbifold is still consistent with modular invariance, so this does not guarantee a nontrivial result.)

As an example, consider T-duality of the boson, which is equivalent to an asymmetric reflection.  Recall that, at the self-dual radius, this reflection is equivalent (by conjugation) to a shift \cite{Ginsparg}, so we can realize a T-duality orbifold of the self-dual compact free boson as our earlier orbifold example with $N=4,\beta=\gamma=1$.  We have already seen that this orbifold acts trivially -- thinking of the projection in the language of (\ref{projstate}), we see that the T-duality operator is generically of order 16 in the twisted sectors.  This is in line with the observations of Harvey and Moore in \cite{HarveyMoore}, where the $g^4$-twisted sector of such an orbifold is shown not to match the untwisted sector, while the $g^{16}$-twisted sector does.

\subsubsection{Two Compact Bosons}
\label{twobosons}

The case of a single boson is limiting.  The power of this flavored formalism is that it readily encompasses both symmetric and asymmetric orbifolds, but as we've just checked, an asymmetric orbifold of a single compact boson is effectively equivalent to a symmetric orbifold.  Luckily, our analysis extends to a collection of bosons.  In order to keep things manageable we'll work with 2, but the generalization to more is straightforward.  

We'll use the same notation as above, now with subscripts distinguishing the bosons.  For instance, the group element now reads $g=g_1g_2=\text{exp}[2\pi i(\frac{\gamma_1 x_1+\beta_1 y_1}{N_1}+\frac{\gamma_2x_2+\beta_2y_2}{N_2})]$.  What twisted sectors should appear in this theory?  Following the line of analysis that led to (\ref{onebosonm}) for the single boson, we find that $m$ should be taken as
\be
m\in\frac{N_1N_2\text{gcd}(N_1,N_2)}{\text{gcd}(N_2^2\beta_1\gamma_1+N_1^2\beta_2\gamma_2,N_1N_2\text{gcd}(N_1,N_2))}\Z_{N_1N_2/\text{gcd}(N_1,N_2)}.
\ee
This is qualitatively the result that we expect -- the orbifold by $g_1g_2$ should indeed have $N_1N_2/\text{gcd}(N_1,N_2)$ sectors.\footnote{For instance, if $g$ is order 2 on the first boson and $g'$ is order 4 on the second, their product should be order 4 and thus we expect 4 total (3 twisted + 1 untwisted) sectors.}

Now that our target space is a two-torus, we can orbifold by asymmetric $\Z_2$ actions and find states satisfying the projection in all sectors, avoiding the empty sectors of the single boson.  As the simplest example, $N_1=N_2=2, \beta_1=\gamma_1=\beta_2=\gamma_2=1$ gives
\begin{multline}
\label{2bosonz2}
|\eta|^{-4}\sum_{\substack{m\in\Z_2 \\ x_1,y_1,x_2,y_2\in\Z^4 \\ x_1+y_1+x_2+y_2=m\text{ mod 2}}}q^{\frac{1}{4}\left[\frac{1}{R_1}\left(x_1-\frac{m}{2}\right)+R_1\left(y_1-\frac{m}{2}\right)\right]^2+\frac{1}{4}\left[\frac{1}{R_2}\left(x_2-\frac{m}{2}\right)+R_2\left(y_2-\frac{m}{2}\right)\right]^2} \\
\bar{q}^{\frac{1}{4}\left[\frac{1}{R_1}\left(x_1-\frac{m}{2}\right)-R_1\left(y_1-\frac{m}{2}\right)\right]^2+\frac{1}{4}\left[\frac{1}{R_2}\left(x_2-\frac{m}{2}\right)-R_2\left(y_2-\frac{m}{2}\right)\right]^2}.
\end{multline}
We can see that this must be generically nontrivial.  In the pre-orbifold theory, the total left and right momenta squared (i.e. $(p_{L,R})_1^2+(p_{L,R})^2_2$) are fractioned in units of lcm$(R_1^2,R_2^2)^{-1}$.  From the orbifold partition function we see that the new squared momenta come in units of lcm$(4R_1^2,4R_2^2)^{-1}$.  However, the resulting partition function is symmetric (this can be seen by taking $y_1\to-y_1+m, y_2\to-y_2+m$).

As a point of comparison, consider Aoki, D'Hoker and Phong \cite{AokiDHokerPhong}.  They construct shift orbifolds by the asymmetric actions $s_R$ and $s_R^2$ which, in our language, correspond to $N=4,\gamma=1,\beta=-1$ and $N=2,\gamma=1,\beta=-1$, respectively.  In the case of orbifolds of multiple bosons by $s_R$ (acting on each), they identify that ``to properly restore the order of $s_R$ to be 4 as it was in the untwisted sector, the dimension of the torus [...] must be divisible by 4.''  This matches our notion that, as the action of $s_R$ is generically of order 16 in the twisted sectors, we would need 4 bosons to guarantee that all sectors remain non-empty.  However, in order to maintain modular invariance, Aoki, D'Hoker and Phong opt for the requirement of group invariance in all sectors.  This means that they consider only multiples of 16 bosons in their $s_R$ orbifolds, and multiples of 4 for $s_R^2$.  While such a requirement guarantees them modular invariance, they miss out on models like (\ref{2bosonz2}) above (which is T-dual to an orbifold by $s_R^2$), which utilize group covariance in twisted sectors.

\subsubsection{Adding a Noncompact Boson}

Let's see what this setup gives when we couple our theory to a noncompact free boson (this will be an example of the fibered CFTs~\cite{Hellerman:2006tx} discussed in section~\ref{subsec:FiberedCFTs}).  This does not obey our initial assumption that the theory had a discrete spectrum, so we'll start the analysis anew.  To begin with, our partition function is
\be
Z=|\eta|^{-2}\int_{-\infty}^\infty dp\text{ }\rho_0(q\bar{q})^\frac{p^2}{2}\text{ Tr}\left[q^{L_0-\frac{c}{24}}\bar{q}^{\bar{L}_0-\frac{\bar{c}}{24}}\right],
\ee
where the trace is over the compact boson Hilbert space, the integral is over the noncompact boson momentum, and where $\rho_0$ is a constant (which is, strictly speaking infinite, as discussed in section~\ref{subsec:ZOrbifold})

We'll take the group element to be
\be
g=e^{2\pi i(Lp+\alpha_LJ_L-\alpha_RJ_R)}
\ee
which generates an infinite-order shift on the noncompact boson and acts as some (possibly asymmetric) translation on the internal theory.  Inserting the group element and doing the integration gives
\be
Z_{0,r}(\tau)=\frac{\rho_0}{\sqrt{\tau_2}}|\eta|^{-2}e^\frac{-\pi L^2r^2}{\tau_2}Z^f(\tau,r\alpha_L,r\alpha_R ).
\ee
Once again, modular transformations give
\be
\label{eq:Zmncross2}
Z_{m,n}(\tau)=\frac{\rho_0}{\sqrt{\tau_2}}|\eta|^{-2}e^\frac{-\pi L^2|n-m\tau|^2}{\tau_2}e^{\pi ik\left(m^2(\tau\alpha_L^2-\bar{\tau}\alpha_R^2)-mn(\alpha_L^2-\alpha_R^2)\right)}Z^f(\tau,(n-m\tau)\alpha_L,(n-m\bar{\tau})\alpha_R).
\ee
where we see a disappearance of any $r$ dependence, as before.  Writing out all of the exponentials and using Poisson resummation on $n$ gives the full partition function as
%\begin{multline}
%Z=\sum_{\substack{m,n\in\Z \\ \text{states } i}}|\eta|^{-2}q^\frac{-c}{24}\bar{q}^\frac{-\bar{c}}{24}\text{exp}\biggl[-\pi\tau_2\biggl(L^2m^2+\frac{1}{L^2}\biggl(n-(\alpha_LJ_L-\alpha_RJ_R)+\frac{km}{2}(\alpha_L^2-\alpha_R^2)\biggr)^2 \\
%-2m(\alpha_LJ_L+\alpha_RJ_R)+m^2k(\alpha_L^2+\alpha_R^2)+2(h_i+\bar{h}_i)\biggr)+2\pi i\tau_1(h_i-\bar{h}_i-mn)\biggr]
%\end{multline}
%in terms of $\tau_1,\tau_2$ and
\begin{multline}
\label{noncompacttimesarbitrary}
|\eta|^{-2}\sum_{\substack{m,n\in\Z \\ \text{states } i}}q^{\frac{1}{4}\left[4h_i-2m(\alpha_LJ_L+\alpha_RJ_R)+m^2k(\alpha_L^2+\alpha_R^2)+m^2L^2+\frac{1}{L^2}\left[n-(\alpha_LJ_L-\alpha_RJ_R)+\frac{km}{2}(\alpha_L^2-\alpha_R^2)\right]^2-2mn\right]-\frac{c}{24}} \\
\times\bar{q}^{\frac{1}{4}\left[4\bar{h}_i-2m(\alpha_LJ_L+\alpha_RJ_R)+m^2k(\alpha_L^2+\alpha_R^2)+m^2L^2+\frac{1}{L^2}\left[n-(\alpha_LJ_L-\alpha_RJ_R)+\frac{km}{2}(\alpha_L^2-\alpha_R^2)\right]^2+2mn\right]-\frac{\bar{c}}{24}}.
\end{multline}
%in terms of $\tau,\bar{\tau}$.
This time around we didn't have to impose any projection.  The fact that we coupled the action on the internal theory with a $\Z$ on the noncompact boson means that all of the sectors can be nonempty.

Additionally, we can look at the limit $L\to 0$ of this theory to make sure it's consistent with our previous results.  Indeed, looking at the $1/L^2$ term and imagining taking $L\to 0$, if its coefficient is nonzero, those states will become infinitely massive and decouple; requiring that the coefficient vanish will give us the projection (\ref{proj}).  The easiest way to see what happens to the overall partition function in this limit is to look at (\ref{eq:Zmncross2}), the partition function prior to resumming on $n$.  Taking $L\to 0$ in that equation (then summing appropriately) clearly gives the noncompact free boson times the $\Z$ orbifold of the internal theory from section~\ref{internaltheory}, up to normalization.  Note the fact that the normalizing constant $\rho_0$ is linear in $L$, so formally the partition function vanishes in this limit.

If we choose to take the compact free boson as our internal theory (with the same choices of $J$ and $\alpha$ as in previous sections), (\ref{noncompacttimesarbitrary}) can be cast in the nicer form
\begin{multline}
\label{noncompacttimesboson}
|\eta|^{-4}\sum_{m,n,x,y\in\Z^4}q^{\frac{1}{4}\left[Lm-\frac{1}{L}\left(n-\frac{x\gamma}{N}-\frac{y\beta}{N}+\frac{m\beta\gamma}{N^2}\right)\right]^2+\frac{1}{4}\left[\frac{1}{R}\left(x-\frac{m\beta}{N}\right)+R\left(y-\frac{m\gamma}{N}\right)\right]^2} \\
\times\bar{q}^{\frac{1}{4}\left[Lm+\frac{1}{L}\left(n-\frac{x\gamma}{N}-\frac{y\beta}{N}+\frac{m\beta\gamma}{N^2}\right)\right]^2+\frac{1}{4}\left[\frac{1}{R}\left(x-\frac{m\beta}{N}\right)-R\left(y-\frac{m\gamma}{N}\right)\right]^2}.
\end{multline}
Compare this to the case investigated in section~\ref{twobosons} where both bosons were initially compact.  The weights in both partition functions take similar forms, but starting with a non-compact boson dramatically simplifies the sum (it comes with absolutely no constraints to worry about).  In fact one can see from this equation exactly how this happens -- instead of needing to introduce a new infinite sum over twisted sectors to handle arbitrary orbifold group elements, the `twisted sectors' from the internal theory's point of view are simply labeled by the now compact secondary boson's momenta, which were already going to be summed over $\Z$ anyways.  The winding of the auxiliary boson then compensates to maintain modular invariance.

%\subsection{Comparison to examples in the literature}

\section{$\Z_N\times\Z_{N'}$ orbifolds}
\label{sec:TwoGenerator}

It is possible to build more complicated groups by iterating the orbifold procedure.  Specifically, any group that can be built by successive extensions by abelian groups (a condition known as solvability) should be amenable to the flavored orbifold method.  Here we sketch a demonstration that the orbifold by $\Z_N$ followed by another by $\Z_{N'}$ is equivalent to an orbifold by $Z_N\times\Z_{N'}$.  We assume that the currents generating these groups commute.

First note that the na\"{i}ve desire would be to begin with our partition function with elements from both groups inserted.  This would give the untwisted sector partial traces $Z_{(0,0),(r,r')}$.  We would hope to make modular transformations to reach the full set of partial traces, $Z_{(m,m'),(n,n')}$.  Unfortunately, not all modular orbits are connected to the untwisted sector, meaning that this procedure necessarily will fail to generate the full partition function.  In order to apply the modular orbit method to this problem we will need to be a little more clever.

The plan, schematically, is to start with $r$ insertions of the first group, $Z_{(0,0),(r,0)}$, and make the usual transformation to $Z_{(j,0),(l,0)}$.  At this point we would usually sum over $j$ and $l$; instead, we will insert elements of the second group into the partial traces, taking us to $Z_{(j,0),(l,r')}$. Can we fill out the modular orbits from here?  A generic SL$(2;\Z)$ transformation given by $(a',b',c',d')$ takes us to $Z_{(a'j-c'l,-c'r'),(d'l-b'j,d'r')}$.  Similarly to before, let us take $r'=\text{gcd}(m',n')$ and choose $c'=-m'/r', d'=n'/r'$.  In order for this transformation to be in SL$(2;\Z)$ we need to choose $a'$ and $b'$ such that $a'n'+b'm'=r'$ -- B\'{e}zout's identity guarantees that we can make such a choice.  Now $(a',b',c',d')$ have all been fixed.  Can we choose $j$ and $l$ in such a way that $a'j-c'l=m$ and $d'l-b'j=n$ for any choice of integers $m,n$?  Write this system of equations in matrix form
\be
\left[\begin{matrix} a' & -c' \\ -b' & d' \end{matrix}\right]\left[\begin{matrix} j \\ l\end{matrix}\right]=\left[\begin{matrix} m \\ n\end{matrix}\right]
\ee
and it becomes plain that, since the coefficient matrix is in SL$(2,\Z)$, we can invert it to get the desired solution.  Picking $j=d'm+c'n, l=b'm+a'n$, we are finally at $Z_{(m,m'),(n,n')}$.

Implementing this procedure for the flavored partition function yields a final orbifold partition function
\be
\label{doubleorbifold}
Z=\sum_{\substack{m\in\Z_N \\ m'\in\Z_{N'} \\ \text{proj.}}}q^{h_i-m\alpha_LJ_L+\frac{km^2}{2}\alpha_L^2-m'\alpha'_LJ'_L+\frac{k'm^{'2}}{2}\alpha^{'2}_L-\frac{c}{24}}\bar{q}^{\bar{h}_i-m\alpha_RJ_R+\frac{km^2}{2}\alpha_R^2-m'\alpha'_RJ'_R+\frac{k'm^{'2}}{2}\alpha^{'2}_R-\frac{\bar{c}}{24}},
\ee
where ``proj.'' in the sum means that all states must simultaneously satisfy the projections
\be
\label{multiproj1}
\alpha_LJ_L-\alpha_RJ_R-\frac{km}{2}(\alpha_L^2-\alpha_R^2)\in\Z
\ee
and
\be
\label{multiproj2}
\alpha'_LJ'_L-\alpha'_RJ'_R-\frac{k'm'}{2}(\alpha^{'2}_L-\alpha^{'2}_R)\in\Z.
\ee
This is exactly what we would obtain beginning from the full $\Z_N$ orbifold (\ref{generalzm}) and performing a $\Z_{N'}$ orbifold on it, as claimed.

Note that, in order to successfully fill out all of the orbits, we were forced insert the elements of the second group into the twisted sector of the first.  While we may have had an (untwisted) action in mind for the second group, it is possible that there are multiple consistent ways it could act in the first group's twisted sectors.  This is related to the phenomenon of discrete torsion.

By way of example, let us look at a $\Z_2\times\Z_2$ orbifold.  For this group there is one orbit disconnected from the untwisted sector, formed from the six partial traces
\be
\label{disconnected}
Z_{\text{disconnected}}=Z_{(0,1),(1,0)}+Z_{(1,0),(0,1)}+Z_{(0,1),(1,1)}+Z_{(1,0),(1,1)}+Z_{(1,1),(0,1)}+Z_{(1,1),(1,0)}.
\ee
The disconnected piece can enter the partition function only as
\be
Z_{\Z_2\times\Z_2\text{ orbifold}}=Z_{\text{connected}}\pm Z_{\text{disconnected}},
\ee
a constraint which can be viewed as arising from modular invariance at higher genus \cite{VafaTorsion}.

When we construct the orbifold as outlined above and insert the second group's elements to arrive at $Z_{(j,0),(l,r')}$, we could just as well insert an additional factor of $e^{\pi i jr'}$.  This would preserve the untwisted action of the second group while consistently modifying it in the $j$-twisted sectors.  After the modular transformation sending us to the full $Z_{(m,m'),(n,n')}$, we find that this modification has become $e^{\pi i(n'm-m'n)}$.  We see immediately that this operator gives $-1$ on the partial traces in the disconnected orbit (\ref{disconnected}), while on the other 10 connected orbits it gives $+1$.

We now consider an example of this behavior in an orbifold of two bosons.  We'll perform a symmetric $\Z_2$ coordinate shift orbifold on each, successively, so the result will be of the form (\ref{doubleorbifold}).  The projections (\ref{multiproj1}) and (\ref{multiproj2}) tell us that $x_1,x_2\in 2\Z$, while $y_1$ and $y_2$ become half integer (shifted respectively by $m/2$ and $m'/2$), bringing us from the theory at $R_1, R_2$ to the theory at $R_1/2, R_2/2$:
\be
|\eta|^{-4}\sum_{x_1,y_1,x_2,y_2}q^{\frac{1}{4}\left[\frac{2}{R_1}x_1+\frac{R_1}{2}y_1\right]^2+\frac{1}{4}\left[\frac{2}{R_2}x_2+\frac{R_2}{2}y_2\right]^2}\bar{q}^{\frac{1}{4}\left[\frac{2}{R_1}x_1-\frac{R_1}{2}y_1\right]^2+\frac{1}{4}\left[\frac{2}{R_2}x_2-\frac{R_2}{2}y_2\right]^2}.
\ee
Now that we've done a $\Z_2\times\Z_2$, having calculated $Z_{(m,m'),(n,n')}$, we can make the other choice for discrete torsion by inserting $e^{\pi i(n'm-m'n)}$ before performing any sums.  The effect of this insertion is to modify and couple the projections: instead of $x_1,x_2\in 2\Z$ we get $x_1+m',x_2+m\in 2\Z$.  Taking into account these constraints, the resulting partition function can be written as
\begin{multline}
|\eta|^{-4}\sum_{s,p,q,r\in\Z^4}q^{\frac{1}{4}\left[\frac{2}{R_1}\left(q+\frac{r}{2}\right)+R_1\left(s-\frac{p}{2}\right)\right]^2+\frac{1}{4}\left[\frac{2}{R_2}\left(s+\frac{p}{2}\right)+R_2\left(q-\frac{r}{2}\right)\right]^2}\\
\times\bar{q}^{\frac{1}{4}\left[\frac{2}{R_1}\left(q+\frac{r}{2}\right)-R_1\left(s-\frac{p}{2}\right)\right]^2+\frac{1}{4}\left[\frac{2}{R_2}\left(s+\frac{p}{2}\right)-R_2\left(q-\frac{r}{2}\right)\right]^2}.
\end{multline}

%\section{Future directions}
%\label{sec:Future}
%
%\subsection{Multiple generators and free groups}
%
%\subsection{Higher genus}
%
%It would be sensible to ask whether these results extend to orbifold theories on higher genus Riemann surfaces.  Our basic method, as laid out in section \ref{sec:proposal}, readily generalizes to the higher genus scenario.  In that situation we would label partial traces by $2g$ group elements, one for each cycle in the homology of the surface.  We would write the analogue of (\ref{eq:ModTransOrbTraces}) for Sp$(2g;\Z)$ and follow the same lines of reasoning.  In future work we intend to lay out the consequences of this procedure.  In particular, with these tools we can begin to address concern (iii) of section \ref{sec:pitfalls} by giving a prescription to directly compute orbifold correlation functions and OPEs from the degeneration of higher genus surfaces.  Again we will find that the flavored case gives us exceptional control over orbifolds by continuous symmetries, and will be able to compute a large number of explicit results at once.

\section{Conclusions and Future Directions}
\label{sec:Conclusion}

We put forth, in section \ref{sec:proposal}, a constructive proposal for defining orbifold partition functions.  Our method emphasizes the role of modular transformations in building these objects and avoids making assumptions about the existence or structure of twisted Hilbert spaces.  Not needing to impose traditional level matching constraints allows us to treat symmetric and asymmetric actions on the same footing -- the phases required to match up the partial traces follow automatically from modular invariance.  Our procedure is guaranteed by construction to produce a modular invariant result, but we have no guarantee a priori that we will end up with a new theory.

In the case of theories with continuous currents we were able to implement our method explicitly, resulting in a completely general expression for the twisted sector partition function, (\ref{generalzm}).  In section \ref{sec:Boson} we showed how this expression captures shift orbifolds of the free boson (or multiple copies thereof), and showed that both mundane and more exotic examples follow from the same framework.

Section \ref{sec:pitfalls} laid out some of the ways our procedure might run into trouble.  One issue was that it is not completely obvious how to generalize our procedure to arbitrary non-cyclic groups.  One potential route would be to lift a finite symmetry group $G$ with $n$ distinct generators to the action of the free group $F_n$ on $n$ generators, with many elements of the free group acting ineffectively on the parent theory.  This is the most direct analog of what we have proposed in the single generator case, where $F_1\cong\Z$.  The free groups have the desirable property that their higher group cohomology vanishes, $H^2(F_n,\U(1))\cong H^3(F_n,\U(1))\cong 0$, so we can again generate partial traces for a $F_n$ orbifold uniquely starting from the untwisted sector partial traces.

When $n>1$ this is a bit confusing, however.  The issue is that we are supposed to include only partial traces $Z_{g,h}$ for which $g$ commutes with $h$.  In $F_n$, however, two elements only commute if they are both powers of a common element, i.e.\ if there exists $k\in F_n$ and integers $p$ and $q$ such that $g=k^p$, $h=k^q$.  These are of course also the only partial traces that we can obtain by modular transformations from the untwisted sector.  Thus many of the partial traces that we would expect from the orbifold by $G$ would simply be missing in the $F_n$ orbifold.

The likely answer is to be found in the previous studies of ineffective group actions~\cite{Pantev:2005rh,Hellerman:2006zs}.  For the cases studied in this paper, we extended a cyclic group to a larger, anomaly free, cyclic group $\widehat{G}$.  The orbifold by the larger group resulted in a direct sum of disconnected identical theories (in the $\widehat{G}\cong\Z$ case; for $\widehat{G}\cong\Z_{KN}$, with $K$ chosen appropriately, only a single theory can be obtained).  In the non-cyclic case, we expect (based on examples in the references above) to again obtain a sum of disconnected theories, but now averaged over possible choices of discrete torsion.  This averaging precisely cancels out the partial traces which are missing from the $F_n$ orbifold, making the whole story consistent.  Another route to examine is the possibility of more general finite anomaly-free extensions, which are proven to exist in~\cite{Wang:2017loc}.  We plan to explore these issues further in subsequent work.

Another natural direction to investigate is the extension to higher genus Riemann surfaces.  Our basic method, as laid out in section \ref{sec:proposal}, readily generalizes to the higher genus scenario.  In that situation we would label partial traces by $2g$ group elements, one for each cycle in the homology of the surface.  We would write the analogue of (\ref{eq:ModTransOrbTraces}) for Sp$(2g;\Z)$ and follow the same lines of reasoning.  In a companion paper (to appear) we will lay out the consequences of this procedure.  In particular, with these tools we can begin to address concern (iii) of section \ref{sec:pitfalls} by giving a prescription to directly compute orbifold correlation functions and OPEs from the degeneration of higher genus surfaces.  Again we find that the flavored case gives us exceptional control over orbifolds by continuous symmetries, and we will be able to compute a large number of explicit results at once.

%We began to address the issue of disconnected modular orbits in section \ref{sec:TwoGenerator} by building an orbifold by a group with multiple generators through an interative process.  
%We hope to address issues of identifying the underlying CFT to which our orbifold partition function belongs in future work by considering orbifolds at higher genus.

\section*{Acknowledgements}

The authors would like to thank O.~Lunin and the other members of the University at Albany string group for helpful conversations.  This material is based upon work supported by the National Science Foundation under Grant No.\ PHY-1820867.  DR would like to thank S.~Hellerman for useful discussion, and would like to thank the Aspen Center for Physics and the Simons Center for Geometry and Physics for hospitality while this work was being completed.

\appendix
\section{Review of path integral formulation of orbifolds}
\label{app:PIFormulation}

Here we review the orbifold construction in the case that a path integral formulation is available for the parent theory, for instance if the parent theory is a theory of free fields $\varphi$ with action $S[\varphi]$.

In that case, the partition function of the parent theory can be written schematically as
\be
Z(\tau,\btau)=\int\mathcal{D}\varphi\,e^{-S_E[\varphi]},
\ee
where we have analytically continued to Euclidean signature (denoted by the subscript $E$), and where the path integral runs over configurations of $\varphi(t_E,x)$ defined on the Euclidean torus with parameter $\tau$, so $\varphi$ satisfies periodic boundary conditions
\be
\varphi(t_E,x)=\varphi(t_E,x+2\pi)=\varphi(t_E+2\pi\tau_2,x+2\pi\tau_1).
\ee
Or, in terms of a complex worldsheet coordinate $z=\tfrac{1}{2\pi}(x+it_E)$,
\be
\varphi(z,\bz)=\varphi(z+1,\bz+1)=\varphi(z+\tau,\bz+\btau).
\ee

In the path integral formulation, the untwisted partial traces are given by
\be
Z_{1,g}(\tau,\btau)=\int\mathcal{D}\varphi_{1,g}e^{-S_E[\varphi]},
\ee
where the subscript on the path integral measure means that we now integrate over fields that satisfy
\be
\varphi(z+1,\bz+1)=\varphi(z,\bz),\quad \varphi(z+\tau,\bz+\btau)=g\cdot\varphi(z,\bz).
\ee

The first step in constructing a modular invariant orbifold partition function is to generalize these partial traces to
\be
\label{eq:ZhgDef}
Z_{h,g}(\tau,\btau)=\int\mathcal{D}\varphi_{h,g}e^{-S_E[\varphi]}=\Tr_{\mathcal{H}_h}\ls\rho_h(g)q^{L_0-\tfrac{c}{24}}\bq^{\bar{L}_0-\tfrac{\bar{c}}{24}}\rs,
\ee
where now the integral is over fields satisfying
\be
\label{eq:hgPeriodicity}
\varphi(z+1,\bz+1)=h\cdot\varphi(z,\bz),\quad \varphi(z+\tau,\bz+\btau)=g\cdot\varphi(z,\bz),
\ee
and $\mathcal{H}_h$ is the Hilbert space built up using fields satisfying the first peridodicity requirement in (\ref{eq:hgPeriodicity}), with $\rho_h$ the representation of the group $G$ acting on this Hilbert space\footnote{As discussed further in section \ref{sec:TwoGenerator}, there can be ambiguity in how this group action is to be defined.}.  Note that these boundary conditions are only consistent if $g$ and $h$ commute, otherwise the result of $\varphi(z+\tau+1,\bz+\btau+1)$ would be ambiguous.  Also, since the action $S_E[\varphi]$ should be invariant under a reflection $z\rr -z$, we have $Z_{h,g}=Z_{h^{-1},g^{-1}}$.

Now in a modular-invariant CFT, $S_E[\varphi]$ will not transform when we act by an $\SL(2,\Z)$ transformation which sends $\tau\rr\tfrac{a\tau+b}{c\tau+d}$.  On the other hand the integration region does transform.  If $\tau'=\tau+1$, then the integration fields appearing in $Z_{h,g}(\tau',\bar{\tau}')$ satisfy
\begin{multline}
g\cdot\varphi(z,\bz)=\varphi(z+\tau',\bz+\bar{\tau}')=\varphi((z+\tau)+1,(\bz+\btau)+1)=h\cdot\varphi(z+\tau,\bz+\btau)\\
\Rightarrow\quad\varphi(z+\tau,\bz+\btau)=(h^{-1}g)\cdot\varphi(z,\bz),
\end{multline}
thus
\be
\mathcal{D}\varphi_{h,g}\longrightarrow\mathcal{D}\varphi_{h,gh^{-1}},\qquad\mathrm{and}\qquad Z_{h,g}(\tau+1,\btau+1)=Z_{h,gh^{-1}}(\tau,\btau).
\ee
Similarly, for $\tau'=-1/\tau$, the two cycles on the torus get exchanged and we find
\be
h\cdot\varphi(z,\bz)=\varphi(z+1,\bz+1)=\varphi(\tfrac{w+\tau}{\tau},\tfrac{\bar{w}+\btau}{\btau})\quad\Rightarrow\quad h\cdot\varphi(w,\bar{w})=\varphi(w+\tau,\bar{w}+\btau),
\ee
\be
g\cdot\varphi(z,\bz)=\varphi(z-\tfrac{1}{\tau},\bz-\tfrac{1}{\btau})=\varphi(\tfrac{w-1}{\tau},\tfrac{\bar{w}-1}{\btau})\quad\Rightarrow\quad g\cdot\varphi(w,\bar{w})=\varphi(w-1,\bar{w}-1),
\ee
where $w=z\tau$ and we have used the conformal invariance of the theory to rescale our coordinates by factors of $\tau$.  Thus we have
\be
Z_{h,g}(-1/\tau,-1/\btau)=Z_{g,h^{-1}}(\tau,\btau).
\ee
Combining these results, we have that
\be
\label{Appeq:ModTransOrbTraces}
Z_{h,g}(\frac{a\tau+b}{c\tau+d},\frac{a\btau+b}{c\btau+d})=Z_{g^{-c}h^a,g^dh^{-b}}(\tau,\btau),\qquad\mathrm{for}\qquad\lp\begin{matrix}a & b \\ c & d\end{matrix}\rp\in\SL(2,\Z).
\ee
Note that the property $gh=hg$ is preserved.

From these considerations, there is an obvious way to construct a modular invariant theory, at least for a finite group $G$.  We simply sum over all possible partial traces.  Thanks to the transformation rule (\ref{eq:ModTransOrbTraces}), this is guaranteed to be modular invariant.  Comparing to the untwisted sector partition function $Z_1$, we deduce that the partial traces should be weighted by a factor of $1/|G|$, so we define the orbifold partition function as
\be
Z_G=\frac{1}{|G|}\sum_{\substack{g,h\in G \\ gh=hg}}Z_{h,g}.
\ee
From the definitions of the partial traces, it is clear that they are invariant under conjugation, since this simply corresponds to a redefinition of the fields $\varphi$,
\be
Z_{h,g}=Z_{khk^{-1},kgk^{-1}},\qquad\forall k\in G.
\ee
Thus for any fixed $h$, the sum $\sum_{g,hg=gh}Z_{h,g}$ depends only on the conjugacy class $C=[h]$ of $h$, and we can define the {\it{twisted sector partition function}}
\be
Z_C=\frac{1}{|G|}\sum_{h\in C}\sum_{\substack{g\in G\\ gh=hg}}Z_{h,g}=\frac{|C|}{|G|}\sum_{\substack{g\in G\\ gh=hg}}Z_{h,g},
\ee
where in the second equality $h$ is any (arbitrary) choice of element in $C$.  Of course we have $Z_1=Z_{[1]}=Z_{\{1\}}$.

We can then rewrite the full orbifold partition function as a sum over twisted sectors.
\be
Z_G=\sum_{C\subset G}Z_C.
\ee

For abelian groups, the conjugacy classes all consist simply of single elements and we have
\be
Z_h:=Z_{\{h\}}=\frac{1}{|G|}\sum_{g\in G}Z_{h,g},\qquad Z_G=\sum_{h\in G}Z_h.
\ee
These will be our primary focus in this paper.

Finally, let us note that when we discuss infinite order groups, we can still use parts of this formalism if we phrase things directly in terms of projection operators rather than using partial traces.  In other words we can define twisted sector partition functions in the operator formalism as
\be
Z_C=\Tr_{\mathcal{H}_h}\ls\Pi_{N_h}q^{L_0-\tfrac{c}{24}}\bq^{\bar{L}_0-\tfrac{\bar{c}}{24}}\rs,
\ee
where $h$ is some arbitrary element in the conjugacy class $C$, $N_h=\{g\in G|gh=hg\}$ is the subgroup of elements which commute with $h$, and $\Pi_{N_h}$ is the projection onto states which are invariant under the action of this subgroup.

%\newpage
%%%%%%%%%%%%%%%%%%%%%%%%%%%%%%%%%%%%%%%%%%%%%%%%%%%%%%%%%%%%

%\bibliographystyle{ieeetr}
%\bibliography{ModularOrbitsPaper}

%\providecommand{\href}[2]{#2}\begingroup\raggedright\begin{thebibliography}{10}

%\end{thebibliography}\endgroup

\end{document}